\documentclass[12pt,aps,preprint]{article}

\usepackage[utf8]{inputenc}

\usepackage{authblk} 
\usepackage{orcidlink}
\usepackage{graphicx,array}
\usepackage{url}
\usepackage{color}
\usepackage{latexsym}
\usepackage{amsthm}
\usepackage{amsmath}
\usepackage{amssymb}
\usepackage{amsfonts}
\usepackage[numbers,sort&compress]{natbib}
\usepackage{bm}
\usepackage{bbm}
\usepackage{slashed}
\usepackage{mathrsfs}
\usepackage{enumerate}
\usepackage{tikz}
\usepackage{siunitx}
\usepackage{mdframed}
\usepackage{setspace}  
\usepackage{esvect}
\usepackage{physics}
\usepackage{enumitem}
\usepackage{booktabs}
\usepackage{tcolorbox}%
\usepackage{fontawesome5}

\numberwithin{equation}{section}

\usepackage{hyperref} 
\hypersetup{
    colorlinks=true,       
    linkcolor=red,          
    citecolor=blue,        
    filecolor=magenta,      
    urlcolor=blue           
}
\usepackage[all]{hypcap} 
\usepackage{multirow}
\usepackage{multicol}

\newcolumntype{M}[1]{>{\centering\arraybackslash}p{#1}}

\usepackage{natbib}
\setlist[description]{leftmargin=\parindent,labelindent=\parindent}

\usepackage[normalem]{ulem} 

\newcommand{\nc}{\newcommand}

\newcommand{\ffa}{f_a}
\newcommand{\fpi}{f_\pi}
\newcommand{\qbar}{\overline{q}}
\newcommand{\mq}{\boldsymbol{m}_{q}}
\newcommand{\cgg}{c_{gg}}
\newcommand{\cQ}{c_{Q}}

\newcommand{\Gsub}{G_{\mu\nu}}
\newcommand{\Gtildesup}{\widetilde{G}^{\mu\nu}}

\newcommand{\cA}{\mathcal{A}}
\newcommand{\bA}{\mathbb{A}}
\newcommand{\bB}{\mathbb{B}}
\newcommand{\cC}{\mathcal{C}}
\newcommand{\diag}{{\rm diag}}
\newcommand{\ks}{\kappa_s}
\newcommand{\ku}{\kappa_u}
\newcommand{\kd}{\kappa_d}

\newcommand{\nl}{\nonumber\\}
\newcommand{\clu}{c_u^L}
\newcommand{\cld}{c_d^L}
\newcommand{\cls}{c_s^L}
\newcommand{\cru}{c_u^R}
\newcommand{\crd}{c_d^R}
\newcommand{\crs}{c_s^R}
\newcommand{\qu}{Q_u}
\newcommand{\qd}{Q_d}
\newcommand{\qs}{Q_s}

\nc{\beq}{\begin{equation}}
\nc{\eeq}{\end{equation}}
\nc{\beqa}{\begin{eqnarray}}  
\nc{\eeqa}{\end{eqnarray}}  
\nc{\bit}{\begin{itemize}}  
\nc{\eit}{\end{itemize}}

\newcommand{\ie}{{\it i.e.}}

\usepackage{floatrow}
\newfloatcommand{capbtabbox}{table}[][\FBwidth]

\usepackage{blindtext}

\allowdisplaybreaks

\title{
\vspace{0 ex}
\begin{flushright}
{\text{\small CPTNP-2025-016}}
\end{flushright} 
\vspace{2ex}
\hfill\\
Wess-Zumino-Witten Interactions of Axions: Three-Flavor }

\author[a,b]{Yang Bai\orcidlink{0000-0002-2957-7319}}
\author[a]{Ting-Kuo Chen\orcidlink{0000-0002-5267-6729}}
\author[c,d]{Jia Liu\orcidlink{0000-0001-7386-0253}}
\author[c]{Xiaolin Ma\orcidlink{0009-0007-1994-9493}}
\affil[a]{\small \it Department of Physics, University of Wisconsin-Madison, Madison, WI 53706, USA } 
\affil[b]{\small \it  HEP Division, Argonne National Laboratory, Argonne, IL 60439, USA} 
\affil[c]{\small \it  School of Physics and State Key Laboratory of Nuclear Physics and Technology, Peking University, Beijing 100871, China} 
\affil[d]{\small \it  Center for High Energy Physics, Peking University, Beijing 100871, China} 

\date{}

\begin{document}

\maketitle
\begin{abstract}
We present a complete Lagrangian describing axion interactions with pseudoscalar and (axial-)vector mesons within the three light-flavor quark framework. This formulation incorporates both the standard chiral Lagrangian and the full Wess-Zumino-Witten (WZW) term. By including instanton effects associated with the anomalous $U(1)_A$ symmetry, we demonstrate that physical observables remain invariant under arbitrary chiral phase rotations of the quark fields. This comprehensive Lagrangian provides a robust and consistent framework for exploring axion phenomenology through its interactions with mesons and gauge bosons. As a demonstration, we compute the decay widths of GeV-scale axions into various mesonic final states for several benchmark axion models.
\end{abstract}

\setlength{\parskip}{0.2ex}

\newpage    
\setcounter{page}{1}  

\begingroup
\hypersetup{linkcolor=black,linktocpage}
\tableofcontents
\endgroup

\newpage

\section{Introduction}\label{sec:intro}

Axions and axion-like particles (ALPs) are pseudo-Nambu–Goldstone bosons arising from the spontaneous breaking of a global U(1) symmetry. They are ubiquitous in theoretical physics, appearing not only as a solution to the strong CP problem in particle physics~\cite{Peccei:1977hh, Peccei:1977ur, Weinberg:1977ma, Wilczek:1977pj}, but also as natural predictions in string theory~\cite{Svrcek:2006yi}. Owing to their non-thermal production in the early universe, axions could constitute all or part of the observed dark matter~\cite{ParticleDataGroup:2022pth}. In order to search for axions/ALPs, it is important to examine their interactions with the Standard Model (SM) particles.

One complication in the study of axion couplings to other particles is the confinement and chiral symmetry breaking nature of low-energy QCD. In this scenario, the axion interactions with quarks and gluons at an ultraviolet (UV) scale must be converted to those with mesons and baryons at an infrared (IR) scale below the QCD scale around 1~GeV. Given the various subtleties in the matching from UV to IR couplings, careful scrutiny of the procedure is required to ensure the chiral-basis independence of the calculations~\cite{Bauer:2021wjo, Ovchynnikov:2025gpx, Bai:2024lpq}. In Ref.~\cite{Georgi:1986df}, the effective chiral Lagrangian with a light axion was derived for the first time, opening the gate to a broader landscape of axion-hadron interactions. Later in Ref.~\cite{Bauer:2021wjo}, the consistent implementation of weak meson decays to axions such as $K^\pm \to \pi^\pm a$ was presented and played the central role in several experimental searches for axions through meson decays~\cite{BaBar:2021ich,Goudzovski:2022vbt,Coloma:2023oxx,HIKE:2023ext,ICARUS:2024oqb}. The main achievement of Ref.~\cite{Bauer:2021wjo}, based on the framework established in Ref.~\cite{Bauer:2020jbp} and now widely adopted in the study of low-energy axion effective field theories (see Refs.~\cite{Blinov:2021say,Bai:2024lpq,Ovchynnikov:2025gpx} for example), is the demonstration of the independence of auxiliary chiral rotation phases of the physical observables such as the effective couplings of the $a \rightarrow \gamma\gamma$ and $K^\pm \to \pi^\pm a$ decay processes. 
Without a careful consistency check for this basis independence of the physical observables, it is rather easy to make a mistake. For instance, Ref.~\cite{Aloni:2018vki} explicitly computed the hadronic decays of axions via axion–neutral-meson mixing, the results of which, however, as pointed out in Ref.~\cite{Ovchynnikov:2025gpx}, turned out to be dependent on the unphysical auxiliary phases and thus unreliable. On the other hand, Ref.~\cite{Ovchynnikov:2025gpx} incorporated the ``pion-shift" method into the Hidden Local Symmetry (HLS) formalism~\cite{Harada:2003jx} to achieve results that are independent of such phases for the interactions they considered, which we will discuss more in Section~\ref{sec:WZW}. 

In addition to the interactions based on the standard chiral Lagrangian, axions could also have topological interactions with hadrons, or Wess–Zumino–Witten (WZW) interactions~\cite{Wess:1971yu,Witten:1983tw}. In our previous work~\cite{Bai:2024lpq}, we presented a \textit{consistent} derivation of the complete WZW interactions of axions by treating the derivative of the axion field as a background gauge field, thereby embedding the axion into the WZW term plus the appropriate counterterm.
The inclusion of WZW counterterm~\cite{Harvey:2007rd, Harvey:2007ca} is necessary for the preservation of fundamental gauge invariance, the cancellation of auxiliary chiral rotation parameters, and the satisfaction of the ‘t Hooft anomaly matching conditions~\cite{tHooft:1979rat}. The study in Ref.~\cite{Bai:2024lpq} was restricted to the two-flavor case with only $u$- and $d$-quarks and omitted the $s$-quark as well as the $\eta$ and $\eta^\prime$ mesons, which significantly limits its phenomenological applicability. In this study, we extend our analysis to the three-flavor case with the aim to provide the complete interactions of axions with mesons for a given UV axion model. 

As we go from the two-flavor to the three-flavor case, one new important issue is how to deal with the anomalous $U(1)_A$ symmetry explicitly broken by the QCD instanton effects~\cite{tHooft:1979rat,Kawarabayashi:1980dp,Kawarabayashi:1980uh}. 
The quantum breaking of $U(1)_A$ prevents the flavor singlet $\eta_0$ from being a massless Goldstone boson and provides a large non-vanishing mass for the $\eta^\prime$ meson in the chiral limit~\cite{Gan:2020aco}.
Therefore, it is essential to account for the QCD $U(1)_A$ anomaly and instanton effects in the effective low-energy theory~\cite{Witten:1980sp, DiVecchia:1980yfw} when it comes to axion interactions with mesons mapped from the UV interactions such as $a G\widetilde{G}$. In the literature, one widely used approach is to rotate away the $a G\widetilde{G}$ term via an axion-dependent chiral phase rotation of the quark fields, at the cost of modifying the quark mass terms and axion–quark couplings. The complete removal of this term requires specific constraints on the rotation phase parameters such as $\Tr(\boldsymbol{\kappa}_q)=1$ (see for example Refs.~\cite{GrillidiCortona:2015jxo,Bauer:2020jbp,Bauer:2021wjo,Bauer:2021mvw,Wang:2024tre,Guo:2025icf} and Section~\ref{sec:Leff} for the details; here $\boldsymbol{\kappa}_q$ denotes the axial-like rotation phases of the quark fields). An alternative strategy is to integrate out the gluonic degrees of freedom and replace them with an effective description in terms of the topological charge density~\cite{DiVecchia:2013swa,Alves:2024dpa,Gao:2024vkw}. In this work, we will examine both approaches in our derivation of the axion couplings, including the WZW interactions, and demonstrate the consistency of the two approaches, which has yet been done explicitly in the literature.  

Following the approach in our previous study~\cite{Bai:2024lpq}, we derive the complete axion interactions with pseudoscalar mesons, vector-like vector mesons, and axial-like vector mesons from both the standard chiral Lagrangian and the full WZW term.
By properly accounting for the QCD $U(1)_A$ anomaly and instanton effects, we will demonstrate that the auxiliary-phase dependence in axion-meson interactions is automatically eliminated for any chirally-rotated bases without imposing any conditions on $\Tr(\boldsymbol{\kappa}_q)$—a result not achieved in the literature. Moreover, our formalism correctly describes the axion interactions that involve both vector and axial-vector fields, which requires a non-trivial cancellation of the $\boldsymbol{\delta}_q$ phases that are the vector-like rotation phases of the quark fields. To understand the dependence of the axion couplings on the various rotation parameters, we will pay special attention to the charge conjugation (C) symmetry and note that axions can be either C-even or C-odd in certain models (see more discussion in Section~\ref{sec:WZW}).
Using three benchmark axion interaction models, we will compute the axion decay patterns and explore some phenomenological implications of the WZW interactions studied in this work.

The structure of this paper is as follows. In Section~\ref{sec:Leff}, we present the axion effective Lagrangians at both the quark- and hadron-levels in the three-flavor framework and explain the matching procedure between them. In Section~\ref{sec:WZW}, we review the WZW formalism developed in Ref.~\cite{Bai:2024lpq}, extend it to the three-flavor case, and demonstrate the cancellation of auxiliary chiral phase parameters in several axion interactions. In Section \ref{sec:decay}, we compute the axion decay widths to various pseudoscalar and (axial-)vector mesons predicted by a few example benchmark axion coupling models. Finally, we conclude our study in Section~\ref{sec:conclusions}.

\section{Axion effective Lagrangian}\label{sec:Leff}

We begin by writing down the axion effective Lagrangian at the quark-level. The effective Lagrangian is defined below the electroweak symmetry breaking scale and we ignore the renormalization-group equation running effects. In our study, we allow charge-conjugation (C) and space-parity (P) violation through the derivative axion-quark couplings, but require CP to be conserved. For the study on CP-violating axions, see Ref.~\cite{Dekens:2022gha} for example. Consequently, 
\begin{equation}
\begin{aligned}
\label{eq:Leff0}
    \mathcal{L}_{\rm eff} &= \mathcal{L}_{\rm SM} + \qbar i\slashed{D}q -\left(\qbar_L \mq q_R + {\rm H.c.}\right) + \frac{1}{2}\partial_\mu a\,\partial^\mu a - \frac{m_{a}^2}{2}a^2  \\
    &\quad + \frac{\partial^\mu a}{\ffa}\left(\qbar_L\gamma^\mu\boldsymbol{k}_Lq_L + \qbar_R\gamma^\mu\boldsymbol{k}_Rq_R\right) + \cgg\frac{\alpha_s}{4\pi}\frac{a}{\ffa}\Gsub\Gtildesup + \frac{a}{\ffa}\sum_{\cA_{1,2}}c_{\cA_1\cA_2}F_{\cA_1\mu\nu}\widetilde{F}^{\mu\nu}_{\cA_2} + \mathcal{L}_c ~,
\end{aligned}
\end{equation}
where $q=(u,d,s)^T$ contains three flavor light quarks, $\mq=\diag(m_u,m_d,m_s)$ is the diagonal and real quark mass matrix, $\cA$ contains $\bA$ (fundamental gauge fields excluding the gluon field, see Eq.~\eqref{eq:A_fields}) and $\bB$ (background (axial-)vector meson fields,\footnote{
The ``background fields'' at the quark level function as fictitious non-dynamical gauge fields associated with the global flavor symmetry and are introduced to track the global anomalies for deriving the necessary counterterms for axion-field-related rotations. Below the QCD scale, they will manifest as the dynamical massive vector mesons. We refer to Refs.~\cite{Bai:2023bbg,Harvey:2007ca,Son:2004tq} for previous studies that also use the background field method.
} see Eq.~\eqref{eq:B_fields}), $F_{\cA}$ denotes the corresponding field strength, the anti-symmetric field strength is defined as $\widetilde{F}_{\mu\nu}=\frac{1}{2}\epsilon^{\mu\nu\rho\sigma}F_{\rho\sigma}$, and $c_{\cA_{1}\cA_{2}}$ denotes the anomalous coupling strength between the axion, $\cA_1$, and $\cA_2$ under the current basis. The covariant derivative $D^\mu$ is
\begin{equation}
    D^\mu = \partial^\mu - i\sum_{\cA}\left(\cA_L^\mu P_L+\cA_R^\mu P_R\right) ,~\quad P_{L,R} = \frac{1\mp\gamma^5}{2} ~,
\end{equation}
with the corresponding (gauge) couplings absorbed into the definitions of $\cA$. Here, we choose the derivative axion-quark couplings to be
\begin{equation}
    \boldsymbol{k}_{L} = \diag(c_u^L,c_d^L,c_s^L) ,~\quad \boldsymbol{k}_{R} = \diag(c_u^R,c_d^R,c_s^R) ~.
\end{equation}
Note that to properly incorporate the axion field, the Goldstone boson associated with the spontaneously broken $U(1)_{\rm PQ}$ symmetry, we shift the background fields~\cite{Bai:2024lpq}
\begin{equation}
    \bB_{L/R}^\mu \to \bB_{L/R}^\mu + \boldsymbol{k}_{L/R}\frac{\partial^\mu a}{\ffa} ~.
    \label{eq:axion-shift}
\end{equation}
The last term in Eq.~\eqref{eq:Leff0}, $\Gamma_c(\bA_{L/R},\bB_{L/R})\equiv\int d^4x\,\mathcal{L}_c(\bA_{L/R},\bB_{L/R})$, is the counterterm needed to cancel the fundamental gauge anomalies after turning on the background vector fields and is given by~\cite{Harvey:2007ca} 
\begin{equation}\label{eq:WZW:counterterm}
\begin{aligned}
    \Gamma_c &= -2\,\cC\bigg\{\int\Tr\left[
        (\bA_Ld\bA_L + d\bA_L\bA_L)\bB_L
        + \frac{1}{2}\bA_L(\bB_Ld\bB_L+d\bB_L\bB_L)\right. \\
        &\quad \left.\hspace{2cm} - \frac{3}{2}i\bA_L^3\bB_L - \frac{3}{4}i\bA_L\bB_L\bA_L\bB_L - \frac{i}{2}\bA_L\bB_L^3
    \right] - (L\leftrightarrow R)\bigg\} ~,
\end{aligned}
\end{equation}
where $\cC=N_c/(48\pi^2)$ with $N_c=3$. We have written down the gauge and background fields in their 1-forms, $\bA(\bB)_{L,R}\equiv \bA(\bB)^\mu_{L,R}dx_\mu$.
Accordingly, $\bA_{L,R}$ are given respectively by
\begin{equation}\label{eq:A_fields}
    \bA_L = \frac{e}{s_w}W^i\boldsymbol{T_i} + \frac{e}{c_w}W^0\boldsymbol{Y}_Q ,~ \bA_R = \frac{e}{c_w}W^0\boldsymbol{Y}_q ~,
\end{equation}
where $W^i$ and $W^0$ stand for the $SU(2)_L$ and $U(1)_Y$ gauge 1-forms, respectively, $s_w\equiv\sin\theta_w$ and $c_w\equiv\cos\theta_w$ with $\theta_w$ as the weak mixing angle, $\boldsymbol{T_i}$ stands for the $SU(2)_L$ generators of the fundamental representation in the $(u,d,s)$ basis and is given by 
\begin{equation}
    \boldsymbol{T_i} = \frac{1}{2}\begin{pmatrix}
        \boldsymbol{\tau}_i & \\
         & -\delta_{i3}
    \end{pmatrix} ,~ i=1,2,3 ~,
\end{equation}
$\boldsymbol{\tau}_i$ representing the Pauli matrices, $\boldsymbol{Y}_Q=\diag(1/6,1/6,1/6)$, and $\boldsymbol{Y}_q=\diag(2/3,-1/3,-1/3)$. 
Note that to properly describe the weak interactions, $c$-quark should also be included, which then allows us to write down a gauge-anomaly-free theory by treating it as a double-copy of the two-flavor scenario that has been studied in Ref.~\cite{Bai:2024lpq}. Once we integrate out the $c$-quark, we will need to include the additionally induced counterterms to ensure the cancellation of the gauge anomalies (see Refs.~\cite{DHoker:1984izu,DHoker:1984mif} for example). For this reason, even though we write down our electroweak gauge 1-forms in the three-flavor basis, we will only consider the weak interactions involving $u$- and $d$-quarks. 
On the other hand, the chiral fields $\bB_{L,R}$ are defined through~\cite{Bai:2024lpq}
\begin{equation}\label{eq:B_fields}
\begin{aligned}
    \bB_V &\equiv \bB_L + \bB_R = g \begin{pmatrix}
        \rho_0+\omega & \sqrt{2}\rho^+ & \sqrt{2}K^{*}_+ \\[1ex]
        \sqrt{2}\rho^- & -\rho_0+\omega & \sqrt{2}K^{*}_0 \\[1ex]
        \sqrt{2}K^{*}_- & \sqrt{2}\,\overline{K}^{*}_0 & \sqrt{2}\phi
    \end{pmatrix} +(\boldsymbol{k}_{L}+\boldsymbol{k}_{R})\frac{da}{\ffa} ~, \\
    \bB_A &\equiv \bB_L - \bB_R = g \begin{pmatrix}
        a_1+f_1 & \sqrt{2}a^+ & \sqrt{2}K^{*}_{A+} \\[1ex]
        \sqrt{2}a^- & -a_1 + f_1 & \sqrt{2}K^{*}_{A0} \\[1ex]
        \sqrt{2}K^{*}_{A-} & \sqrt{2}\,\overline{K}^{*}_{A0} & \sqrt{2}f_s
    \end{pmatrix} +(\boldsymbol{k}_{L}-\boldsymbol{k}_{R})\frac{da}{\ffa} ~,
\end{aligned}
\end{equation}
where $g\approx\sqrt{12\pi}$~\cite{ParticleDataGroup:2022pth}. Note that above the QCD and below the electroweak scale, the meson background fields serve as fictitious background gauge fields used to keep track of the global-global-gauge and global-global-global anomalies and only become dynamic below the QCD scale~\cite{Witten:1980sp,DiVecchia:1980yfw,DiVecchia:2013swa,Bai:2024lpq,Harvey:2007rd}.

Below the QCD scale, the QCD degrees of freedom will change from quarks and gluons to hadrons as a result of confinement and chiral symmetry breaking. While one can capture the dynamics of mesons using the chiral Lagrangian described by the pion Goldstone fields, it is also essential to include the QCD $U(1)_A$ anomaly and instanton effect in the low-energy theory~\cite{Witten:1980sp,DiVecchia:1980yfw}, which is sensitive to the treatment of the $aG\widetilde{G}$ term. In the literature, one common way is to perform a chiral rotation on the quark fields to remove this term while modifying the mass and axion-quark coupling terms (see Ref.~\cite{GrillidiCortona:2015jxo} for example). Another way is to integrate out the gluon degrees of freedom and describe the effective theory using the topological charge density (see Ref.~\cite{DiVecchia:2013swa} for example). However, to our knowledge, no literature has ever addressed the consistency between the physical axion couplings derived from these two approaches, which we now address.

Following the prescription of Refs.~\cite{Gao:2024vkw,Alves:2024dpa,tHooft:1986ooh}, we match the quark-level Lagrangian to the $U(3)$ chiral Lagrangian including the QCD $U(1)_A$ anomaly and instanton effect. We first define the topological susceptibility as~\cite{Kaiser:2000gs}
\begin{equation}
    \tau \equiv \lim_{q\to0}\left[-i\int d^4x\,e^{iqx}\langle0\vert \mbox{T}\left\{\frac{\alpha_s}{8\pi}\Gsub\Gtildesup(x)\,\frac{\alpha_s}{8\pi}\Gsub\Gtildesup(0)\right\} \vert0\rangle \right] ~,
\end{equation}
where $\mbox{T}$ stands for time ordering. In the large $N_c$ limit with $N_f$ flavors of massless quarks, the Witten-Veneziano relation~\cite{Witten:1979vv,Veneziano:1979ec} gives that
\begin{equation}
    m_0^2 = \frac{4\,N_f}{F_0^2}\tau_{\rm GD} \approx \frac{4\,N_f}{F_0^2}\tau ~,
\end{equation}
where $N_f=3$, $\tau=\tau_{\rm GD}+\text{(quark-loop corrections)}$ with $\tau_{\rm GD}$ being the component of $\tau$ resulting purely from gluon dynamics, $m_0$ is the mass of the singlet pseudo-Goldstone field $\eta_0$ associated with the spontaneously broken $U(1)_A$ symmetry, and $F_0$ is the corresponding decay constant. Here we approximate $\tau\approx\tau_{\rm GD}$ as a result of the leading-order large $N_c$ approximation. As we match the quark-level Lagrangan to the chiral Lagrangian below the QCD scale, we will take the leading-order result $F_0=\fpi\approx130$~MeV, the pion decay constant, while matching the instanton effect~\cite{Rosenzweig:1979ay}
\begin{equation}
    \overline{\theta}\frac{\alpha_s}{8\pi}\Gsub\Gtildesup \to -\frac{\tau}{2} \left(-i\log\det U-\overline{\theta} \right)^2 ~,
\end{equation}
where $\overline{\theta}\equiv\theta+\arg\det(M_dM_u)$ is the effective strong CP phase, $M_{u,d}$ being the mass matrices of the up- and down-type quarks, respectively. For the purpose of our study, we assume $\overline{\theta}=0$, whether it is induced by the axion relaxation mechanism or not. Accordingly, one has
\begin{equation}
    U = \exp\left[(\sqrt{2}i/\fpi)\pi^a\boldsymbol{t}^a\right]\equiv \exp[(\sqrt{2}i/\fpi)\boldsymbol{\Phi}] ~,
\end{equation}
which describes the nonet (pseudo-)Goldstone fields of the $U(3)_L\times U(3)_R \rightarrow U(3)_V$ symmetry and $\boldsymbol{t}^a$ (with $\Tr(\boldsymbol{t}^a\boldsymbol{t}^b)=2\delta^{ab}$) denotes the generators of the broken $U(3)_A$, or more explicitly, 
\begin{equation}
    \boldsymbol{\Phi} = \begin{pmatrix}
        \pi_0 + \frac{1}{\sqrt{3}}\eta_8 + \sqrt{\frac{2}{3}}\eta_0 & \sqrt{2}\pi^+ & \sqrt{2}K^+ \\[1ex]
        \sqrt{2}\pi^- & -\pi_0 + \frac{1}{\sqrt{3}}\eta_8 + \sqrt{\frac{2}{3}}\eta_0 & \sqrt{2}K^0 \\[1ex]
        \sqrt{2}K^- & \sqrt{2}\,\overline{K}^0 & -\frac{2}{\sqrt{3}}\eta_8 + \sqrt{\frac{2}{3}}\eta_0
    \end{pmatrix} ~.
\end{equation} 
Consequently, the singlet $\eta_0$ mass $m_0$ is given via $\tau=f_\pi^2 m_0^2/12$. After combining this term with the axion-gluon interaction, one has
\begin{align}\label{eq:axial-anomaly}
    \mathcal{L}_{\chi{\rm PT}} \supset \displaystyle -\frac{\tau}{2}\left(-i\log\det U-2\cgg\frac{a}{\ffa}\right)^2=-\frac{m_0^2}{2}\left(\eta_0-\frac{\cgg}{\sqrt{3}}\frac{f_\pi}{\ffa}a\right)^2 ~,
\end{align}
where we have implicitly used the relation $\det [\exp X]= \exp(\Tr [X])$. Note that Eq.~\eqref{eq:axial-anomaly} matches the results in Refs.~\cite{Gao:2024vkw,Alves:2024dpa} after replacing $c_{gg}/\ffa\to-1/(2\ffa)$~\cite{Bauer:2017ris,Bauer:2020jbp}. A direct consequence of this term is the mass mixing between $\eta_0$ and $a$, as can be seen in the mass matrix given later in Eq.~\eqref{eq:mass_matrix}. Later in this study, we will fix the value of $m_0$ through the physical neutral pseudoscalar meson masses.

To the leading order in $\mathcal{O}(f_\pi/\ffa)$, axion only mixes with the pseudoscalar mesons through this term $\propto m_0^2$ at every order of the isospin-breaking parameter $\delta\equiv (m_d-m_u)/(m_u+m_d)$, which could serve as a consistency check for the calculation of the axion-meson mixing.~\footnote{
For instance, the term given in Eq.~(B.1c) of Ref.~\cite{Blinov:2021say} is not proportional to $m_0^2$, which signals an error in the axion-meson mixing calculations.}
Up to two-derivative terms in the chiral Lagrangian, we arrive at
\begin{equation}
\label{eq:LXPT}
\begin{aligned}
    \mathcal{L}_{\chi{\rm PT}} &= \frac{f_\pi^2}{8}\Tr\left[\left(D^\mu U\right)\left(D_\mu U\right)^\dagger\right] + \frac{f_\pi^2}{4}B_0\Tr\left[\mq U^\dagger + {\rm H.c.}\right] -\frac{m_0^2}{2}\left(\eta_0-\frac{\cgg}{\sqrt{3}}\frac{f_\pi}{\ffa}a\right)^2 \\
    &\quad + \frac{1}{2}(\partial_\mu a)(\partial^\mu a) - \frac{m_{a}^2}{2}a^2 + \frac{a}{\ffa}\sum_{\cA_{1,2}}c_{\cA_1\cA_2}F_{\cA_1\mu\nu}\widetilde{F}^{\mu\nu}_{\cA_2} ~,
\end{aligned}
\end{equation}
where the pion matrix covariant derivative is given by
\begin{equation}
    D^\mu U = \partial^\mu U - i\sum_{\cA}\left[\cA_L^\mu U-U\cA_R^\mu\right] ~,
\end{equation}
and $B_0\approx m_\pi^2/(m_u+m_d)$. 

It is worth noting that the axion derivative coupling to quarks are contained in the $\cA_{L/R}$, see the background field shift in Eq.~\eqref{eq:axion-shift}. The original axion-gluon interaction $a G \widetilde{G}$ is not eliminated by chiral rotation, but integrated out and leads to $m_0^2$ term in Eq.~\eqref{eq:LXPT}.
Here we have dropped $\mathcal{L}_c$ and will include it in the full WZW term to be introduced in Section~\ref{sec:WZW}. We summarize the properties of the vector, axial-vector, and pseudoscalar mesons mentioned in this study in Table~\ref{tab:mesons}.
\begin{table}[ht!]
\centering
\caption{Ground-state vector, axial-vector, and neutral pseudoscalar mesons with $u$, $d$, $s$ content from PDG~\cite{ParticleDataGroup:2022pth}. Note that for charged mesons, they do not have a fixed charge-conjugation quantum number. Also, $K^*_A(1270)$ refers to $K^*_1(1270)$ and $f_s(1420)$ to $f_1(1420)$ in PDG.}
\label{tab:mesons}
\renewcommand{\arraystretch}{1.3}
\small
\setlength{\tabcolsep}{4pt}
\begin{tabular}{ccccc}
\toprule
Category & Particle & Quark Content & Mass (MeV) & G-parity \\
\midrule
\multirow{7}{*}{\shortstack[c]{Vector \\ ($J^{PC}=1^{--}$)}} 
 & $\rho^0(770)$ & $\frac{u\bar{u} - d\bar{d}}{\sqrt{2}}$ & 775 & +1 \\
 & $\rho^\pm(770)$ & $u\bar{d}/d\bar{u}$ & 775 & +1 \\
 & $\omega(782)$ & $\frac{u\bar{u} + d\bar{d}}{\sqrt{2}}$ & 783 & -1 \\
 & $\phi(1020)$ & $s\bar{s}$ & 1019 & -1 \\
 & $K^{*}_{\pm}(892)$ & $u\bar{s}/s\bar{u}$ & 892 & — \\
 & $K^{*}_0(892)$ & $d\bar{s}$ & 896 & — \\
 & $\overline{K}^{*}_0(892)$ & $s\bar{d}$ & 896 & — \\
\midrule
\multirow{7}{*}{\shortstack[c]{Axial-vector \\ ($J^{PC}=1^{++}$)}} 
 & $a_1^0(1260)$ & $\frac{u\bar{u} - d\bar{d}}{\sqrt{2}}$ & 1230 & -1 \\
 & $a_1^\pm(1260)$ & $u\bar{d}/d\bar{u}$ & 1230 & -1 \\
 & $f_1(1285)$ & $\frac{u\bar{u} + d\bar{d} + s\bar{s}}{\sqrt{3}}$ & 1285 & +1 \\
 & $K_{A\pm}^{*}(1270)$ & $u\bar{s}/s\bar{u}$ & 1270 & — \\
 & $K_{A0}^{*}(1270)$ & $d\bar{s} $ & 1270 & — \\
 & $\overline{K}_{A0}^{*}(1270)$ & $ s\bar{d}$ & 1270 & — \\
 & $f_s(1420)$ & $s\bar{s}$ & 1426 & +1 \\
\midrule
\multirow{7}{*}{\shortstack[c]{Pseudoscalar \\ ($J^{PC}=0^{-+}$)}}
 & $\pi^0$ & $\frac{u\bar{u} - d\bar{d}}{\sqrt{2}}$ & 135 & -1 \\
 & $\eta$ & $\frac{u\bar{u} + d\bar{d} - 2s\bar{s}}{\sqrt{6}}$ & 548 & +1 \\
 & $\eta'$ & $\frac{u\bar{u} + d\bar{d} + s\bar{s}}{\sqrt{3}}$ & 958 & +1 \\
 & $K^0$ & $d\bar{s}$ & 498 & — \\
 & $\overline{K}^0$ & $s\bar{d}$ & 498 & — \\
 & $\pi^\pm$ & $u\bar{d}/d\bar{u}$ & 140 & — \\
 & $K^\pm$ & $u\bar{s}/s\bar{u}$ & 494 & — \\
\bottomrule
\end{tabular}%
\renewcommand{\arraystretch}{1}
\end{table}

On the other hand, one can consider a chiral axion-dependent phase rotation on the quark fields as 
\begin{equation}
\label{eq:chiral}
    q \to \exp\left[-i\cgg\left(\boldsymbol{\delta}_{q}+\boldsymbol{\kappa}_{q}\gamma^5\right)\frac{a}{\ffa}\right]q ~,
\end{equation}
where the auxiliary and unphysical parameters $\boldsymbol{\delta}_{q}=\diag(\delta_u,\delta_d,\delta_s)$ and $\boldsymbol{\kappa}_{q}=\diag(\kappa_u,\kappa_d,\kappa_s)$, which gives 
\begin{equation}\label{eq:Leffp}
    \mathcal{L}_{\rm eff} = \mathcal{L}_{\rm eff}(q,\cA_{L/R}\vert\mq,\boldsymbol{k}_{L/R},\cgg) \to \mathcal{L}_{\rm eff}(q,\cA_{L/R}\vert\mq^\prime,\boldsymbol{k}_{L/R}^\prime,\cgg^\prime) + \delta\mathcal{L}_a^{\rm ano} ~.
\end{equation}
As we pointed out earlier, one common treatment is to choose $\Tr(\boldsymbol{\kappa}_q)=1$ to get rid of the axion-gluon interaction. However, we do not impose this condition and will show later that all physical quantities are automatically independent of the rotation parameters regardless of our choice of $\Tr(\boldsymbol{\kappa}_q)$ as long as we properly treat the $U(1)_A$ anomaly and instanton effect [see Eq.~\eqref{eq:axial-anomaly}].
Denoting $\boldsymbol{\theta}_{L/R}\equiv\boldsymbol{\delta}_{q}\mp\boldsymbol{\kappa}_{q}$, one has the axion-dependent mass matrix
\begin{equation}\label{eq:mqprime}
    \mq^\prime(a) = \exp(i\cgg\boldsymbol{\theta}_{L}\frac{a}{\ffa})\mq\exp(-i\cgg\boldsymbol{\theta}_{R}\frac{a}{\ffa}) ~,
\end{equation}
the modified derivative axion-quark couplings
\begin{equation}\label{eq:kprime}
    \boldsymbol{k}^\prime_{L/R}(a) = \exp\left(i\cgg\boldsymbol{\theta}_{L/R}\frac{a}{\ffa}\right)\left(\boldsymbol{k}_{L/R}+ \cgg\boldsymbol{\theta}_{L/R}\right)\exp\left(-i\cgg\boldsymbol{\theta}_{L/R}\frac{a}{\ffa}\right) ~,
\end{equation}
the modified anomalous axion-gluon coupling,
\begin{equation}\label{eq:cggprime}
    \cgg^\prime = \cgg[1-\Tr(\boldsymbol{\kappa}_q)] ~,
\end{equation}
and the variation of the anomalous axion couplings
\begin{equation}\label{eq:Lano}
    \delta\mathcal{L}_a^{\rm ano} = - \delta\left[\mathcal{L}_{\rm WZW} + \mathcal{L}_c\right]\left(\cgg\boldsymbol{\theta}_{L},\cgg\boldsymbol{\theta}_{R}\right) ~,
\end{equation}
where $\Gamma_{\rm WZW}(U,\cA_L,\cA_R)\equiv\int d^4x\mathcal{L}_{\rm WZW}(U,\cA_L,\cA_R)$ is the WZW term that we will introduce in Section~\ref{sec:WZW}. We note that the anomaly induced above the QCD scale should match to that below the QCD scale by the WZW term as a result of~'t Hooft anomaly matching \cite{tHooft:1979rat}, and thus the notation ``WZW''. Accordingly, we can match the rotated axion effective Lagrangian to the chiral Lagrangian as
\begin{equation}
    \mathcal{L}_{\chi{\rm PT}} = \mathcal{L}_{\chi{\rm PT}}(U,\cA_{L/R}\vert\mq,\boldsymbol{k}_{L/R},\cgg) \to \mathcal{L}_{\chi{\rm PT}}(U,\cA_{L/R}\vert\mq^\prime ,\boldsymbol{k}_{L/R}^\prime,\cgg^\prime) + \delta\mathcal{L}_{\rm WZW}^{\rm ano} ~,
\end{equation}
where $\delta\mathcal{L}_{\rm WZW}^{\rm ano}=\delta\mathcal{L}_{a}^{\rm ano}$, which we will discuss in the next section.

\section{WZW interactions of axions}\label{sec:WZW}

\begin{figure}[th!]
    \centering
    \includegraphics[width=0.99\linewidth]{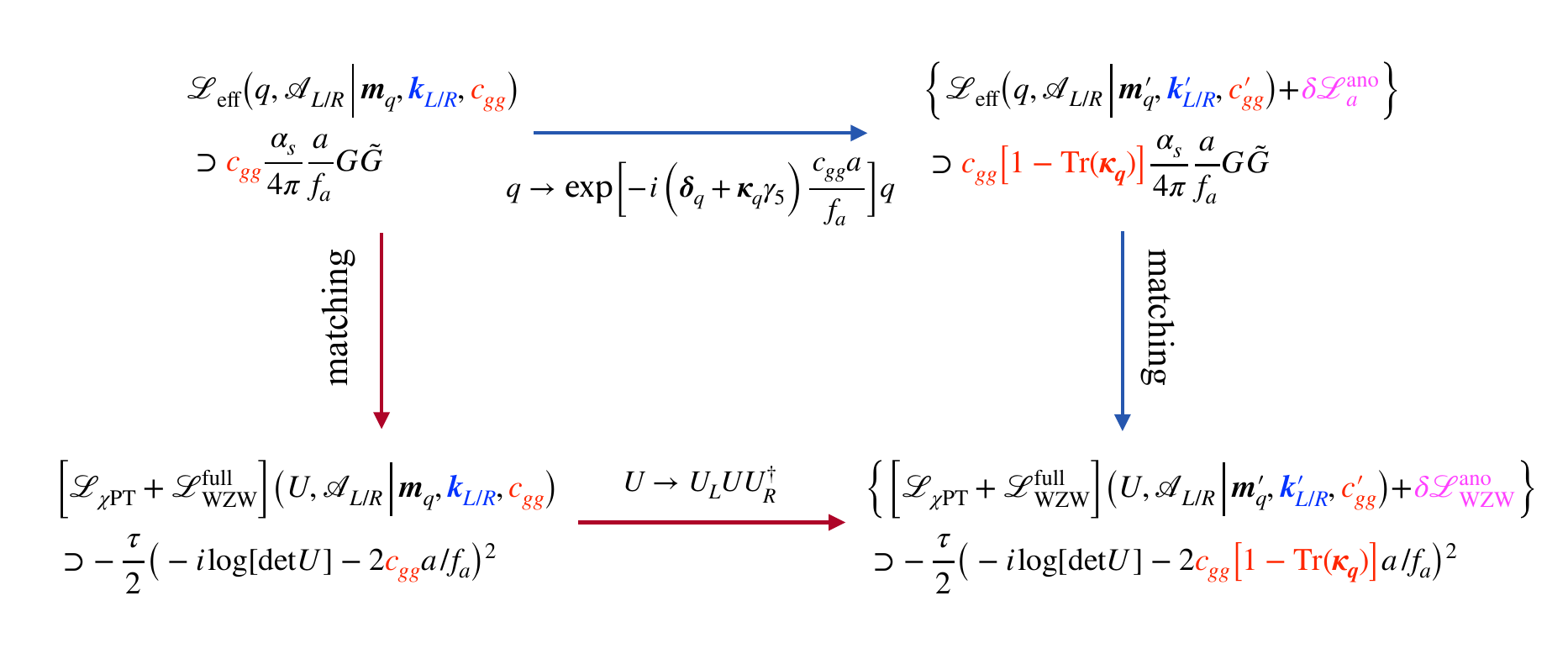}
    \caption{Diagrammatic illustration of the relations among the Lagrangians defined in Eqs.~\eqref{eq:Leff0} (upper left), \eqref{eq:Leffp} (upper right), \eqref{eq:Lfull0} (lower left), and \eqref{eq:Lfullp} (lower right). The definitions of the primed quantities on the right-hand side are given in Eqs.~\eqref{eq:mqprime} ($\boldsymbol{m}_q^\prime$), \eqref{eq:kprime} ($\mathbf{k}_{L/R}^\prime$), and \eqref{eq:cggprime} ($\cgg^\prime$), while $\delta\mathcal{L}_a^{\rm ano}=\delta\mathcal{L}_{\rm WZW}^{\rm ano}$ is defined via Eqs.~\eqref{eq:Lano} and \eqref{eq:variation}.
    }
    \label{fig:bases}
\end{figure}

The WZW term~\cite{Wess:1971yu,Witten:1983tw} is a crucial component of the chiral Lagrangian that describes the anomaly structure of QCD below the QCD scale. By incorporating the (axial-)vector fundamental and meson gauge fields into the WZW term, the global-global-global and global-global-gauge anomalies at the quark-level are reproduced as a result of~'t Hooft anomaly matching~\cite{Harvey:2007ca}. As we pointed out in Ref.~\cite{Bai:2024lpq}, it is essential to embed the axion into the WZW term to produce consistent amplitudes for physical processes. The full WZW term is given by~\cite{Harvey:2007ca}
\begin{align}\label{eq:WZW}
	\Gamma_{\rm WZW}^{\rm full}(U,\cA_L,\cA_R) &= \int d^4x\mathcal{L}_{\rm WZW}^{\rm full} \nonumber\\
        &= \Gamma_0(
 U) + \mathcal{C}\int\Tr\Big\{ (\cA_L\alpha^3+\cA_R\beta^3) - \frac{i}{2}[(\cA_L\alpha)^2-(\cA_R\beta)^2] \nonumber\\
	&\quad +i(\cA_LU\cA_R^\dagger\alpha^2-\cA_RU^\dagger\cA_LU\beta^2) + i(d\cA_RdU^\dagger\cA_LU-d\cA_LdU\cA_RU^\dagger) \nonumber\\
	&\quad + i[(d\cA_L\cA_L+\cA_Ld\cA_L)\alpha + (d\cA_R\cA_R+\cA_Rd\cA_R)\beta] \nonumber\\
	&\quad + (\cA_L^3\alpha+\cA_R^3\beta) - (d\cA_L\cA_L+\cA_Ld\cA_L)U\cA_RU^\dagger \nonumber\\
	&\quad + (d\cA_R\cA_R+\cA_Rd\cA_R)U^\dagger\cA_LU + (\cA_LU\cA_RU^\dagger\cA_L\alpha + \cA_RU^\dagger\cA_LU\cA_R\beta) \nonumber\\
	&\quad + i\Big[\cA_L^3U\cA_RU^\dagger - \cA_R^3U^\dagger\cA_LU - \frac{1}{2}(U\cA_RU^\dagger\cA_L)^2\Big] \Big\} + \Gamma_c ~,
\end{align}
where $\Gamma_c$ is given in Eq.~\eqref{eq:WZW:counterterm}, $\alpha=dUU^\dagger$, $\beta=U^\dagger dU$, and
\begin{equation}\label{eq:Gamma_0}
	\Gamma_0(U) = -\frac{i\,\mathcal{C}}{5}\int d^5x\epsilon^{ABCDE}\Tr(\alpha_A\alpha_B\alpha_C\alpha_D\alpha_E) ~,
\end{equation}
with $A,\cdots,E=0,1,2,3,4$. We note again that the variation of $\Gamma_{\rm WZW}^{\rm full}$ with respect to the chiral rotation given in Eq.~\eqref{eq:chiral:U} [or equivalently Eq.~\eqref{eq:chiral}] matches that at the quark-level, both of which are given in Eq.~\eqref{eq:variation}.
Combined with the chiral Lagrangian given in Eq.~\eqref{eq:LXPT},  the full axion chiral Lagrangian is then given by
\begin{eqnarray}\label{eq:Lfull0}
    \mathcal{L}_{\rm axion}^{\rm full} = \left[\mathcal{L}_{\chi{\rm PT}}+\mathcal{L}_{\rm WZW}^{\rm full}\right](U,\cA_{L/R}\vert\mq,\boldsymbol{k}_{L/R},\cgg) ~,
\end{eqnarray}
which, after the chiral rotation, becomes
\begin{equation}\label{eq:Lfullp}
    \mathcal{L}_{\rm axion}^{\rm full} \to (\mathcal{L}_{\rm axion}^{\rm full})^\prime = \left[\mathcal{L}_{\chi{\rm PT}}+\mathcal{L}_{\rm WZW}^{\rm full}\right](U,\cA_{L/R}\vert\mq^\prime,\boldsymbol{k}_{L/R}^\prime,\cgg^\prime) + \delta\mathcal{L}_{\rm WZW}^{\rm ano} ~.
\end{equation}
We summarize the relations among the four bases given in Eqs.~\eqref{eq:Leff0}, \eqref{eq:Leffp}, \eqref{eq:Lfull0}, \eqref{eq:Lfullp} in Figure~\ref{fig:bases}. 

Explicitly, for the axion-dependent chiral rotation at meson level, one can use the transformation rules given in Ref.~\cite{Harvey:2007ca},
\begin{equation}\label{eq:chiral:U}
      U \to  U_L U U_R^\dagger ,~ U_{L/R}=e^{-i\cgg\boldsymbol{\theta}_{L/R}\frac{a}{\ffa}} ,~ \mathbb{A}_{L/R} \to \mathbb{A}_{L/R} ,~ \mathbb{B}_{L/R} \to \mathbb{B}_{L/R} + \cgg\boldsymbol{\theta}_{L/R}\frac{da}{\ffa} ~,
\end{equation}
which then leads to 
\begin{align}
     &\Gamma_{\rm WZW}\left( U , \mathcal{A}_L, \mathcal{A}_R \right)+\Gamma_c \left(\mathbb{A}_{L/R}, \mathbb{B}_{L/R}\right)\nonumber\\
     &\quad \to \Gamma_{\rm WZW}\left(U, \mathcal{A}_L, \mathcal{A}_R \right)+\Gamma_c \left(\mathbb{A}_{L/R}, \mathbb{B}_{L/R}\right)-\delta\left[\Gamma_{\rm WZW}+\Gamma_c\right](\cgg\boldsymbol{\theta}_{L},\cgg\boldsymbol{\theta}_{R})\nonumber \\
     &\quad\quad\equiv \Gamma_{\rm WZW}\left(U, \mathcal{A}_L, \mathcal{A}_R \right)+\Gamma_c \left(\mathbb{A}_{L/R}, \mathbb{B}_{L/R}\right) + \int d^4x\,\delta \mathcal{L}^{\rm ano}_{\rm WZW} ~,
     \label{eq:after_trans}
\end{align}
where $\delta\mathcal{L}_{\rm WZW}^{\rm ano}=\delta\mathcal{L}_a^{\rm ano}$ comes from both the variation of the UV anomalous axion couplings and the full WZW term [see Eq.~\eqref{eq:WZW}], and
\begin{equation}\label{eq:variation}
\begin{aligned}
    \delta\left[\Gamma_{\rm WZW} + \Gamma_c\right]\left(\boldsymbol{\theta}_{L},\boldsymbol{\theta}_{R}\right) &= -2\,\cC\frac{a}{\ffa}\int\Tr\bigg\{
        \boldsymbol{\theta}_L\bigg[
            3(d\bA_L-i\bA_L^2)^2 + 3(d\bA_L-i\bA_L^2)(D\bB_L) + D\bB_LD\bB_L \\
        &\quad - \frac{i}{2}D(\bB_L)^3 + i\bB_L(d\bA_L-i\bA_L^2)\bB_L - (d\bA_L-i\bA_L^2)\bB_L^2\bigg]
    - (L\leftrightarrow R)\bigg\}  ~,
\end{aligned}
\end{equation}
with $D\bB_{L,R}=d\bB_{L,R}-i\bA_{L,R}\bB_{L,R}-i\bB_{L,R}\bA_{L,R}$ and $d\bA-i\bA^2$ is the covariant field strength. 

Compared to Ref.~\cite{Aloni:2018vki}, their framework misses both the anomalous and the full WZW contributions, leading to unphysical phase-dependent results for the axion decay amplitudes. It is also worth noting that the transformation in Eq.~\eqref{eq:chiral:U} is equivalent to shifting $\mathbf{k}_{L/R}\to\mathbf{k}^\prime_{L/R}$ [see Eq.~\eqref{eq:kprime}]. We have explicitly checked that this transformation relation is also consistent with the method of ``pion shift'' in Ref.~\cite{Ovchynnikov:2025gpx}, which uses the framework of hidden local symmetry (HLS)~\cite{Bando:1984ej,Fujiwara:1984mp,Bando:1987br} and produces identical results to the ones derived in this study for the axion couplings to gauge and vector-like vector-meson ($J^{PC}=1^{--}$) fields (see also Ref.~\cite{Geng:2025fuc} for a related study on the SM mesons). In our study, however, we have also included the description of the axion coupling to axial-like vector-meson ($J^{PC}=1^{++}$) field interactions, which is not achieved in the framework of Ref.~\cite{Ovchynnikov:2025gpx} and further requires the independence check of the vector-like $\boldsymbol{\delta}_q$ rotation phases. Moreover, the complete WZW interactions induced by the axion-quark couplings also require a proper embedding of the related terms into $\mathcal{L}_{\rm WZW}^{\rm full}$, which is also lacking in Ref.~\cite{Ovchynnikov:2025gpx}. In principal, one can further incorporate these interactions into the HLS framework by including all fourteen counterterms in the generalized HLS framework~\cite{Kaiser:1990yf}, though we find it more straightforward to use our framework based on Ref.~\cite{Harvey:2007ca}.

In the following, we discuss the mixing between the axion and the neutral mesons, which is essential to the consistency of the physical quantities manifested by their independence of the auxiliary phases, and then demonstrate the phase cancellation of several example interactions.

\subsection{Axion-meson mixing}\label{subsec:mixing}

In general, axion can mix with both pseudoscalar and axial-vector mesons. However, since its mixing with axial-vector mesons lacks the resonant enhancement at the meson mass poles~\cite{Ovchynnikov:2025gpx}, we only consider its mixing with pseudoscalar mesons, which does exhibit resonant enhancement and is therefore more important.
Here we present the most general mixing pattern that involves all axion-gluon and axion-quark interactions.
As a leading approximation that offers a great simplification, we follow Ref.~\cite{Aloni:2018vki,Ovchynnikov:2025gpx} to fix the $\eta$-$\eta^\prime$ mixing angle $\theta_{\eta\eta^\prime}$ such that $\sin\theta_{\eta\eta^\prime}=-1/3$ and $\cos\theta_{\eta\eta^\prime}= 2\sqrt{2}/3$, which is equivalent to choosing
\begin{equation}\label{eq:replace}
    \delta = \frac{m_d-m_u}{m_d+m_u} ,~ m_d = (1+\delta)\frac{m_\pi^2}{2m_\eta^2-m_\pi^2}m_s ,~ m_{\eta^\prime}^2 = 4m_\eta^2 - 3m_\pi^2 ,~ m_K^2 = m_\eta^2 - \frac{\delta\,m_\pi^2}{2} ~.
\end{equation}
Along with Eq.~\eqref{eq:Lfullp}, we can then derive the kinetic and mass mixing matrices of the axion and neutral pseudoscalar mesons defined by
\begin{equation}
    \mathcal{L}_{\rm axion}^{\rm full} \supset \frac{1}{2}(\partial^\mu P^T) Z (\partial_\mu P) - \frac{1}{2}P^T M^2 P ,~ P = (a,\pi^0,\eta,\eta^\prime)^T ~,
\end{equation}
where
\begin{equation}
    Z = \begin{pmatrix}
        1 & K_{a\pi^0} & K_{a\eta}&K_{a\eta^\prime}\\[1ex]
       K_{a\pi^0} & 1 & 0 & 0 \\[1ex]
       K_{a\eta} & 0 & 1 & 0 \\[1ex]
    K_{a\eta^\prime} & 0 & 0 & 1
    \end{pmatrix} ~,
\end{equation}
\begin{equation}\label{eq:mass_matrix}
    M^2 = \begin{pmatrix}
        m_{a}^2 & M_{a\pi^0}^2 & M_{a\eta}^2 & M_{a\eta^\prime}^2 \\[1ex]
        M_{a\pi^0}^2 & m_\pi^2 & -\sqrt{\frac{2}{3}}\delta m_\pi^2 & -\frac{\delta m_\pi^2}{\sqrt{3}} \\[1ex]
        M_{a\eta}^2 & -\sqrt{\frac{2}{3}}\delta m_\pi^2 & m_\eta^2 & 0 \\[1ex]
        M_{a\eta^\prime}^2 & -\frac{\delta m_\pi^2}{\sqrt{3}} & 0 & m^2_{\eta^\prime}
    \end{pmatrix} ~,
\end{equation}
with
\begin{align}
K_{a\pi^0} &= \epsilon\frac{c^L_d-c^R_d-c^L_u+c^R_u-(2\kd-2\ku)c_{gg}}{2\sqrt{2}} ~,\nonumber\\
K_{a\eta} &= -\epsilon\frac{c^L_d-c^L_s+c^L_u-c^R_d+c^R_s-c^R_u+(2\ks-2\kd-2\ku)c_{gg}}{2\sqrt{3}} ~,\nonumber\\
K_{a\eta^\prime} &= -\epsilon \frac{c^L_d+2c^L_s+c^L_u-c^R_d-2c^R_s-c^R_u-(2\ku+2\kd+4\ks)c_{gg}}{2\sqrt{6}} ~,\nonumber\\
M_{a\pi^0}^2 &= \epsilon c_{gg}\frac{m_\pi^2[(1-\delta)\kappa_u-(1+\delta)\kappa_d]}{\sqrt{2}} ~,\nonumber\\
M_{a\eta}^2 &= \epsilon c_{gg}\frac{m_\pi^2[1+\delta(\kappa_d-\kappa_u)]+m_\eta^2(-1+\kappa_d-\kappa_s+\ku)}{\sqrt{3}} ~,\nonumber\\
M_{a\eta^\prime}^2 &= \epsilon c_{gg}\frac{4m_\eta^2(-1+\kd+2\ks+\ku)+m_\pi^2[4+(-3+\delta)\kd-6\ks-3\ku-\delta \ku]}{\sqrt{6}} ~,
\end{align}
and $\epsilon = f_\pi/f_a$.
Note that in the leading-order chiral perturbation theory, we cannot reproduce both the physical $\eta$ and $\eta^\prime$ masses. To properly incorporate the correct mass poles while ensuring auxiliary phase independence, we have to make the replacement $m_\eta^2 \to (m_{\eta^\prime}^2 + 3m_\pi^2)/4$ whenever the $\eta^\prime$ field is invoked during the perturbation process, which gives
\begin{equation}
\begin{aligned}
    a &= a_{\rm phys} - \sum_{P^0=\pi^0,\eta,\eta^\prime}h(P^0,m_{P^0})P^0_{\rm phys} ~, \\
    P^0 &= P^0_{\rm phys} - \sum_{P^{0\prime}\neq P^0} \frac{M^2_{P^0P^{0\prime}}}{m^2_{P^0}-m^2_{P^{0\prime}}}P^{0\prime}_{\rm phys} + h(P^0,m_a)a_{\rm phys} ~,
\end{aligned}
\end{equation}
where
\begin{equation}
    h(P^0,m_X) = \frac{1}{m_a^2-m_{P^0}^2}\left[
        M_{aP^0}^2 - m_X^2K_{aP^0} + \sum_{P^{0\prime}\neq P^0}M^2_{P^0P^{0\prime}} \frac{M^2_{aP^{0\prime}}-m_X^2K_{aP^{0\prime}}}{m_X^2-m_{P^{0\prime}}^2}
    \right] ~.
\end{equation}
To simplify the notations for the following discussion, we denote $\theta_{P^0a}\equiv h(P^0,m_a)$, which are given explicitly by
\begin{align}\label{eq:mixing_angles}
    \theta_{\pi^0 a} = \frac{\epsilon}{2\sqrt{2}(m_\pi^2-m_a^2)} &\big\{ m_a^2\left[c_d^L-c_u^L-c^R_d+c^R_u-(2\kd-2\ku)c_{gg}\right] \nonumber\\
    &\quad +2m_\pi^2c_{gg}\left[(1+\delta)\kd+(-1+\delta)\ku\right]\big\} ~, \nonumber\\
    \theta_{\eta a} = -\frac{\epsilon}{2\sqrt{3}(m_\eta^2-m_a^2)}&\left\{m_a^2\left[c_d^L-c_s^L+c_u^L-c_d^R+c_s^R-c_u^R-(2\kd-2\ks+2\ku)c_{gg}\right]\right\} \nonumber\\
    -\frac{2\epsilon}{2\sqrt{3}(m_\eta^2-m_a^2)} &\left\{m_\pi^2\left[1+\delta(\kd-\ku)\right]c_{gg}+m_\eta^2(-1+\kd-\ks+\ku)c_{gg}\right\} ~, \nonumber\\
    \theta_{\eta^\prime a} = -\frac{\epsilon}{2\sqrt{6}(m_{\eta'}^2-m_a^2)}&\left\{m_a^2\left[c_d^L+2c_s^L+c_u^L-c_d^R-2c_s^R-c_u^R-(2\kd+4\ks+2\ku)c_{gg}\right]\right\} \nonumber\\
    -\frac{2\epsilon}{2\sqrt{6}(m_{\eta'}^2-m_a^2)}&\big\{(m_{\eta'}^2+3m_\pi^2)(-1+\kd+2\ks+\ku)c_{gg} \nonumber \\
    &\quad -m_\pi^2\left[-4-(-3+\delta)\kd+6\ks+3\ku+\delta\ku\right]c_{gg}\big\} ~.
\end{align}

Note that in the literature, one common choice is set $\boldsymbol{\kappa}_q\propto \mq^{-1}$ to eliminate the mass mixing terms (for instance, see Ref.~\cite{Aloni:2018vki}). In the recent Ref.~\cite{Ovchynnikov:2025gpx}, they have performed a more complete demonstration of phase independence starting from the basis where the axion-gluon coupling was first eliminated by taking $\Tr(\boldsymbol{\kappa}_q)=1$. In this study, we show that with the $U(1)_A$ anomaly term included [see Eq.~\eqref{eq:axial-anomaly}], the independence of the auxiliary phase is manifested in any basis, even without the requirement of $\Tr(\boldsymbol{\kappa}_q)=1$.

\subsection{Auxiliary phase independence of physical observables}\label{subsec:kappa}

In Ref.~\cite{Bai:2024lpq}, we have demonstrated the phase independence of several effective couplings in the two-flavor scenario after first rotating out the $aG\widetilde{G}$ interaction using Eq.~\eqref{eq:chiral} with $\Tr(\boldsymbol{\kappa}_q)=1$.  In the three-flavor scenario, however, things become more complicated. In addition to the new contributions from the strange quark and mesons, there is also the complication coming from the QCD instanton effect. Nevertheless, as we demonstrate below, taking all of these into consideration allows us to show the phase-independence of the physical couplings in any basis, which has not been achieved in the previous literature~\cite{Bai:2024lpq,Ovchynnikov:2025gpx,Bauer:2020jbp}.

A typical axion decay process derived from the model $\mathcal{L}_{\rm axion}^{\rm full}$ given in Eq.~\eqref{eq:Lfull0} contains three components: the contact interactions from the chiral Lagrangian $\mathcal{L}_{\rm \chi PT}$, the corresponding pseudoscalar meson decay process plus axion-meson mixing, and the full WZW interactions from $\mathcal{L}_{\rm WZW}^{\rm full}$ given in Eq.~\eqref{eq:WZW}. Note that if one performs a chiral rotation on the physical basis and arrives at the corresponding model $(\mathcal{L}_{\rm axion}^{\rm full})^\prime$ given in Eq.~\eqref{eq:Lfullp}, the third component will further include the anomalous interactions $\delta\mathcal{L}_{\rm WZW}^{\rm ano}$ whose details are given in Eq.~\eqref{eq:variation}. In the following, we demonstrate the auxiliary phase independence of three example axion decay processes: $a\to\gamma\gamma$, $a\to\omega\gamma$, and $a\to f_1\gamma$.

\subsubsection{$a\to \gamma\gamma$}\label{subsubsec:gammagamma}

The matrix element of the $a\to\gamma\gamma$ decay is given by
\begin{align}
    i\mathcal{M}(a\to\gamma\gamma) &= \left<\gamma\gamma|\epsilon_{\mu\nu\rho\sigma}\partial^\mu (\gamma)^\nu\partial^{\rho}(\gamma)^\sigma|a\right>\times\left[c_{\rm WZW}+c_{\rm ano}+\sum_{p=\pi^0,\eta,\eta'}c_{p}\theta_{pa}\right] \nl 
    &\equiv \left<\gamma\gamma|\epsilon_{\mu\nu\rho\sigma}\partial^\mu (\gamma)^\nu\partial^{\rho}(\gamma)^\sigma|a\right> \times c_{\gamma\gamma}^{\rm eff} ~,
\end{align}
where
\begin{align}\label{eq:Ceff_gamgam_components}
c_{\rm WZW} &= 0 ,~~
c_{\rm ano} = -\cgg\frac{e^2 N_c \left(  Q_d^2 \kappa_d +  Q_s^2 \kappa_s + Q_u^2 \kappa_u \right)}{4 f_a \pi^2} ,~~
c_{\pi^0} = \frac{e^2 N_c \left( Q_d^2 - Q_u^2 \right)}{4 \sqrt{2} f_{\pi} \pi^2} ~, \nl
~~c_{\eta_8} &= -\frac{e^2 N_c \left( Q_d^2 - 2 Q_s^2 + Q_u^2 \right)}{4 \sqrt{6} f_{\pi} \pi^2} ,~~ c_{\eta_0} = -\frac{e^2 N_c \left( Q_d^2 + Q_s^2 + Q_u^2 \right)}{4 \sqrt{3} f_{\pi} \pi^2} ~, \nl
c_{\eta} & =\frac{2\sqrt{2}}{3}c_{\eta_8}+\frac{1}{3}c_{\eta_0} ,~~~c_{\eta'}=-\frac{1}{3}c_{\eta_8}+\frac{2\sqrt{2}}{3}c_{\eta_0} ~,
\end{align}
while the mixing angles are given in Eq.~\eqref{eq:mixing_angles}. Note that $c_{\rm WZW}=0$ because $\gamma$ is a fundamental vector-like gauge field, as we pointed out in Ref.~\cite{Bai:2024lpq}. After summing up all the terms using the mixing angles in Eq.~\eqref{eq:mixing_angles} that are also $\kappa_q$-dependent, one finds that all the auxiliary phases cancel \textit{without} needing to impose any condition on $\Tr (\boldsymbol{\kappa}_q)$. Consequently, the physical effective coupling is given by

\begin{align}
c_{\gamma\gamma}^{\rm eff} &= 
\frac{c_{gg} e^2 N_c}{24\pi^2 \ffa (m_a^2 - m_\eta^2)(m_a^2 - m_{\eta'}^2)}
\begin{aligned}[t]
    \bigg\{&2 m_{\eta'}^2 m_\pi^2 (\qd^2 - \qs^2 + \qu^2) \\
    &- m_a^2 \left[
        3 m_\pi^2 (\qd^2 + \qu^2) 
        - 2 m_\eta^2 (\qd^2 - \qs^2 + \qu^2)-m_{\eta'}^2(\qd^2+2\qs^2+\qu^2)
    \right] \\
    &- m_{\eta}^2 \left[
        3 m_{\eta'}^2 (\qd^2 + \qu^2)
        - m_\pi^2 (\qd^2 + 2 \qs^2 + \qu^2)
    \right]\bigg\}
\end{aligned}
\nl[1.5ex]
&+ \frac{c_{gg} e^2 N_c (\qd^2 - \qu^2) \, \delta m_\pi^2}
{24\pi^2 \ffa (m_a^2 - m_\eta^2)(m_a^2 - m_{\eta'}^2)(m_a^2 - m_\pi^2)}
\begin{aligned}[t]
    \bigg\{&2 m_{\eta'}^2 m_\pi^2 + m_\eta^2 (-3m_{\eta'}^2 + m_\pi^2)\\
    &+ m_a^2 (2m_\eta^2 + m_{\eta'}^2 - 3m_\pi^2) \bigg\}
\end{aligned}
\nl[1.5ex]
&- \frac{e^2 N_c m_a^2}{48 \pi^2 \ffa}
\begin{aligned}[t]
   \bigg\{ &\frac{3 (\cld - \clu - \crd + \cru)(Q_d^2 - Q_u^2)}{m_a^2 - m_\pi^2} \\
    &- \frac{2 (\cld - \cls + \clu - \crd + \crs - \cru)(Q_d^2 - Q_s^2 + Q_u^2)}{m_\eta^2 - m_a^2} \\
    &- \frac{(\cld + 2\cls + \clu - \crd - 2\crs - \cru)(Q_d^2 + 2 Q_s^2 + Q_u^2)}{m_{\eta'}^2 - m_a^2}\bigg\}
\end{aligned}
\nl[1.5ex]
&- \frac{e^2 N_c \, \delta m_a^2 \, m_\pi^2}
{24\pi^2 \ffa (m_a^2 - m_\eta^2)(m_a^2 - m_{\eta'}^2)(m_a^2 - m_\pi^2)}
\begin{aligned}[t]
    \bigg\{&(3m_a^2 - m_\eta^2 - 2m_{\eta'}^2)
    \left[(\cld - \crd) Q_d^2 - (\clu - \cru) Q_u^2\right] \\
    &- (m_\eta^2 - m_{\eta'}^2)
    \big[
        (\cls - \crs) Q_d^2 
        + (\cld - \crd) Q_s^2 \\
    &\qquad\qquad
        - (\clu - \cru) Q_s^2 
        - (\cls - \crs) Q_u^2
    \big]\bigg\} ~.
\end{aligned}
\end{align}

As a consistency check, we keep only the $c_{gg}$-terms and take the $m_{\eta,\eta^\prime}\to\infty$ limit to retrieve the following two-flavor result,
\begin{align}
    c_{\gamma\gamma}^{\rm eff}=-\frac{e^2N_c}{8\pi^2 f_a}\bigg[(Q_d^2+Q_u^2)+\frac{m_\pi^2(Q_d^2-Q_u^2)}{m_a^2-m_\pi^2}\bigg] ~,
\end{align}
which exactly matches the previous results in Refs.~\cite{Bai:2024lpq,Bauer:2020jbp}.

It is worth mentioning that $c_{\gamma\gamma}^{\rm eff}$ only depends on the combination of $c^L_q-c^R_q$ for $q=u,d,s$ through the mixing of axion with the pseudoscalar mesons as a result of C-conservation in QCD. On the other hand, the WZW term does not contribute to this process [see Eq.~\eqref{eq:Ceff_gamgam_components}] while the anomalous $\cgg$-terms, which are dependent on the $\kappa$-phases, drop out of the final expression as expected. One can further generalize this conclusion to other axion decay channels. For simplicity, let us only consider the photon field in the fundamental gauge sector. Consequently, the full WZW term will only contain interactions of either the form $d\gamma\wedge \bB_{A}\wedge \bB_{V}$ or $d\bB_{V} \wedge \bB_{A}\wedge \bB_{V}$.
Here, we first note that for a general combination of $(c^L_q,c^R_q)$, C-symmetry is violated, while in the scenario with only non-zero $c^L_q-c^R_q$ ($c^L_q+c^R_q$) coupling, axion can be assigned to be C-even (C-odd) for C-symmetry conservation. Another way to put this is to separate the C-even $c^L_q-c^R_q$ and C-odd $c^L_q+c^R_q$ components of the derivative axion-quark couplings in the language of spurions, which we stick to in the remainder of the study. For the decays to two vector-like vector mesons or one vector-like vector meson plus one photon, the axion field can only arise from the C-even $\bB_A$ term ($\propto c^L_q-c^R_q$) in the two types of WZW interactions mentioned previously. Along with the C-conserving meson mixing and that the unphysical $\kappa$-dependent anomalous terms must drop out of the final result, one can conclude that the associated effective couplings can only depend on $c^L_q-c^R_q$. As for the decays to one vector-like vector meson/photon plus one axial-like vector meson, there is no contribution from the meson mixing due to C-symmetry in QCD, while the axion can only emerge from the C-odd $\bB_V$ components ($\propto c^L_q+c^R_q$) in the two types of WZW interactions mentioned above, and thus the associated effective couplings must depend on $c^L_q+c^R_q$. We will verify these statements explicitly in the following two case studies on $a\to\omega\gamma$ and $a\to f_1\gamma$.

\subsubsection{$a\to \omega\gamma$}

We repeat the same exercise for the $a\to\omega\gamma$ decay, whose
effective coupling is given by
\begin{equation}
    c_{\omega\gamma}^{\rm eff} = c_{\rm WZW}+c_{\rm ano}+\sum_{p=\pi^0,\eta,\eta'}c_{p}\theta_{pa}  ~,
\end{equation}
where
\begin{align}
c_{\rm WZW} &= \frac{egN_c(c_d^LQ_d-c_d^RQ_d+c_u^LQ_u-c_u^RQ_u-2Q_dc_{gg}\kappa_d-2Q_uc_{gg}\ku)}{16\pi^{2}\ffa} ~, \nl
c_{\rm ano} &= -\frac{egN_c(Q_d\kd c_{gg}+Q_u\ku c_{gg})}{8\pi^2\ffa} ~,\nl
c_{\pi^0} &= \frac{egN_c(Q_d-Q_u)}{4\sqrt{2}f_\pi \pi^2},~~c_{\eta_8}=-\frac{egN_c(Q_d+Q_u)}{4\sqrt{6}f_\pi \pi^2},~~c_{\eta_0}=-\frac{egN_c(Q_d+Q_u)}{4\sqrt{3}f_\pi \pi^2} ~, \nl
c_{\eta} &= \frac{2\sqrt{2}}{3}c_{\eta_8}+\frac{1}{3}c_{\eta_0},~~~c_{\eta'}=-\frac{1}{3}c_{\eta_8}+\frac{2\sqrt{2}}{3}c_{\eta_0} ~.
\end{align}
One can check that the auxiliary phases indeed still cancel \textit{without} needing to impose any condition on $\Tr (\boldsymbol{\kappa}_q)$. Consequently, the physical coupling is given by
\begin{align}
    c_{\omega\gamma}^{\rm eff} &= \frac{c_{gg}egN_c(Q_d+Q_u)}{24\pi^2 \ffa (m_a^2-m_\eta^2)(m_a^2-m_{\eta'}^2)}
    \begin{aligned}[t]
   \bigg\{ & m_a^2(2m_\eta^2+m_{\eta'}^2-3m_\pi^2)\\
    &+m_\eta^2(m_\pi^2-3m_{\eta'}^2)+2m_{\eta'}^2m_\pi^2\bigg\}
\end{aligned}
\nl[1.5ex]
    & + \frac{\cgg egN_c(Q_d-Q_u)\delta m_\pi^2}{24\pi^2 \ffa(m_a^2-m_\pi^2)(m_a^2-m_\eta^2)(m_a^2-m_{\eta'}^2)} 
\begin{aligned}[t]
   \bigg\{ &m_a^2(2m_\eta^2+m_{\eta'}^2-3m_\pi^2) \\
    &+m_\eta^2(m_\pi^2-3m_{\eta'}^2)+2m_{\eta'}^2m_\pi^2\bigg\}
\end{aligned}
\nl[1.5ex]
    &+\frac{egN_c}{48\pi^2 \ffa} 
\begin{aligned}[t]
   \bigg\{ &3Q_d(c_d^L-c_d^R)+3Q_u(c_u^L-c_u^R)-\frac{3m_a^2(Q_d-Q_u)(c_d^L-c_d^R-c_u^L+c_u^R)}{m_a^2-m_\pi^2} \\
    & \frac{2m_a^2(Q_d+Q_u)(c_u^L-c_u^R+c_d^L-c_d^R+\crs-\cls)}{m_\eta^2-m_a^2}\\
    &\frac{m_a^2(Q_d+Q_u)(\clu-\cru+\cld-\crd+2\cls-2\crs)}{m_{\eta'}^2-m_a^2}\bigg\}
\end{aligned}
\nl[1.5ex]
    &-\frac{egN_c\delta m_a^2m_\pi^2}{24\pi^2 \ffa(m_a^2-m_\pi^2)(m_a^2-m_\eta^2)(m_a^2-m_{\eta'}^2)} 
\begin{aligned}[t]
   \bigg\{ &(-3m_a^2+m_\eta^2+2m_{\eta'}^2)\left[Q_u(\clu-\cru)\right. \\
    &\left.-Q_d(\cld-\crd)\right]-(m_\eta^2-m_{\eta'}^2)(Q_d-Q_u)(\cls-\crs)\bigg\}~.
\end{aligned}
\end{align}
Note that the result only depends on $c^L_q-c^R_q$, as discussed in Section~\ref{subsubsec:gammagamma}.

\subsubsection{$a\to f_1 \gamma$}\label{subsubsec:f1gamma}

Before diving into the algebraic details, we first make a few comments on this interaction. Note that this interaction can only be induced by the C-odd axion couplings while the $aG\widetilde{G}$ term and meson interactions only by the C-even couplings, and thus there is no contribution from the axion-meson mixing to this interaction. Furthermore, since the terms that include exactly one $\gamma$ and one axial-like vector meson $\bB_A$ in $\mathcal{L}_{\rm WZW}^{\rm full}$ all take the form $d\gamma\wedge \bB_{V}\wedge \bB_{A}$, the non-vanishing WZW contribution to the $af_1\gamma$ decay process, or $c_{\rm WZW}$, comes only from the C-odd axion-quark interactions, \ie, the interactions that involve $(\boldsymbol{k}_L+\boldsymbol{k}_R)da/\ffa\subset\bB_V$. Similar results also apply to $c_{\rm ano}$. Finally, in contrast to previous examples where only vector-like vector mesons/gauge fields are involved, the auxiliary $\boldsymbol{\delta}_q$ phases become relevant in this interaction. Consequently, one has
\begin{align}
    c_{f_1\gamma}^{\rm eff} = c_{\rm WZW}+c_{\rm ano}+\sum_{p=\pi^0,\eta,\eta'}c_{p}\theta_{pa} ~,
\end{align}
where
\begin{align}
c_{\rm wzw}&=\frac{e g N_c \left( \cld Q_d + \crd Q_d + \clu Q_u + \cru Q_u + 2 Q_d \delta_d + 2 Q_u \delta_u \right)}{16 \pi^2\ffa } ~, \nl
c_{\rm ano}&=-\frac{e g N_c \left( Q_d \delta_d + Q_u \delta_u \right)}{8 \pi^2\ffa} ~,\nl
c_{\pi^0}&=0,~~c_{\eta_8}=0,~~c_{\eta_0}=0 ~,
\end{align}
which gives 
\begin{align}
    c_{f_1\gamma}^{\rm eff}=\frac{egN_c[(c_d^L+c_d^R)Q_d+(c_u^L+\cru)Q_u]}{16\pi^{2}f_a} ~.
\end{align}
Note that the $\delta_{u,d}$ phases also cancel without any conditions imposed. Moreover, the effective coupling depends only on $c^L_q+c^R_q$, again confirming the discussion in Section~\ref{subsubsec:gammagamma}. For the minimal QCD axion, which only has the $c_{gg}$ coupling, this decay channel vanishes, according to the C-symmetry argument stated above.

\subsubsection{Additional comment on $Z$ boson interactions}

Here we briefly comment on the WZW interactions involving the $Z$ boson.~\footnote{While one can certainly integrate out the $Z$ boson field at the meson energy level, keeping it dynamical is helpful for the study of neutrino-meson interactions such as that conducted in Ref.~\cite{Harvey:2007rd}.} Since $Z$ contains both vector and axial-vector components, its interactions with the other particles in general violate C-symmetry. As a result, one cannot apply the previous argument in Section~\ref{subsubsec:f1gamma} to analyze the allowed WZW interactions of the $Z$ boson, and it is also crucial to take into account both the $\boldsymbol{\delta}_q$ and $\boldsymbol{\kappa}_q$ phases when checking the physical consistency of the corresponding effective couplings. The corresponding couplings involving the axion and the $Z$ boson can be found in the \texttt{Mathematica} notebook on \href{https://github.com/nun3366/Axion-WZW-3}{\color{black}{\faGithub\,\underline{nun3366/Axion-WZW-3}}} which we will introduce later.   

\section{Axion decays}\label{sec:decay}

To properly calculate the decay widths of the axion, we follow Ref.~\cite{Aloni:2018vki} to incorporate a data-driven form factor function $\mathcal{F}(m_a)$ fitted from the $e^+e^-$ data in Refs.~\cite{BaBar:2004ytv,BaBar:2007ceh,BaBar:2017zmc} to account for the higher-order corrections to the effective axion couplings derived previously based only on the leading-order chiral Lagrangian. We also truncate the calculation up to $m_a=2$~GeV since chiral perturbation theory is, in principle, only reliable up to the scale of $\sim 4\pi f_\pi\approx1.7$~GeV, while the integration of the $\mathcal{F}$ function should allow us to push the limit further up to $2$~GeV (note that Ref.~\cite{Aloni:2018vki} performs the calculations up to $m_a=3$~GeV).

For the possible axion decay channels, we mainly follow the framework laid out in Refs.~\cite{Aloni:2018vki,Ovchynnikov:2025gpx,Cheng:2021kjg} but replace the WZW couplings with the correct/complete ones derived in this work. We also note that Ref.~\cite{Ovchynnikov:2025gpx} has provided an open source \texttt{Mathematica} notebook with useful numerical frameworks, based on which we have built a new notebook with the complete WZW interactions integrated in. We have also provided details of the WZW interaction derivations in the notebook, which is available on \href{https://github.com/nun3366/Axion-WZW-3}{\color{black}{\faGithub\,\underline{nun3366/Axion-WZW-3}}}. In the following, we briefly introduce the axion decay channels considered and provide the combined decay width plots of a few different benchmark models. Note that we ignore the decay widths of the final-state mesons and only considered finite-width approximations, and thus the results involving large-width mesons such as $a_1$ in the final state should be taken only as an approximation.

\subsection{$a\to 2V$ channels}

The effective vertex of an axion coupling to two spin-1 fields comes from the WZW interactions and can only take up the Lorentz structure of $S^{\rm vx}[\cA_1\cA_2]C_{\cA_1\cA_2}\mathcal{F}(m_a)\epsilon_{\mu\nu\rho\sigma}a\partial^\mu \cA_1^\nu\partial^\rho \cA_2^\sigma$, with $\cA_{1,2}$ denoting the spin-1 gauge bosons/mesons and $C_{\cA_1\cA_2}$ the effective coupling strength of the interaction. Here, $S^{\rm vx}$ denotes the symmetry factor, where $S^{\rm vx}=2$ if $\cA_1=\cA_2$ and $S^{\rm vx}=1$ if otherwise. We have also inserted the form factor $\mathcal{F}$ into the effective vertex. The associated decay width is given by 
\begin{align}
    \Gamma_{a\to \cA_1\cA_2}=S^{\rm ps}[\cA_1\cA_2](S^{vx})^2
    \left|C_{\cA_1\cA_2}\right|^2\frac{m_a^3}{32\pi}\left[1+\frac{(m_{\cA_1}^2-m_{\cA_2}^2)^2}{m_a^4}-\frac{2(m_{\cA_1}^2+m_{\cA_2}^2)}{m_a^2}\right]^{3/2}\mathcal{F}(m_a)^2 ~,
\end{align}
with $S^{\rm ps}$ denoting phase space symmetry factor for identical particles, where $S^{\rm ps}=1/2$ if $\cA_1=\cA_2$ and $S^{\rm ps}=1/2$ if otherwise. As discussed previously, in the presence of C-odd axion couplings, $\cA_{1,2}$ can be a mixture of vector and axial-vector fields. We further note that although C- and P-symmetries can be violated individually through the simultaneous presence of the C-even and C-odd derivative axion-quark couplings, CP remains intact and the axion is a CP-odd state in the current framework. As a result, effective vertices of the form $aF_{\mu\nu}F^{\mu\nu}$ will not show up in our model, and such terms can arise only in the presence of CP-violating axions (see Section~\ref{sec:Leff} and Ref.~\cite{Dekens:2022gha}), which is beyond the scope of this study.

\subsection{$a\to 4P$ channels}
The contact interaction between axion and four pseudoscalar mesons is absent from the standard chiral Lagrangian due to the spurious pion-number parity $(-1)^{N_p}$~\cite{Witten:1983tw,Kaymakcalan:1983qq,Harvey:2007rd}, $N_p$ being the total number of pseudoscalar mesons and axion (consider the mixing between axion and pseudoscalar mesons). Therefore, this interaction can only come from other terms, such as the WZW term and the interactions involving scalar and tensor mesons, the latter of which have been discussed in Refs.~\cite{Ovchynnikov:2025gpx,Aloni:2018vki}. For the WZW term, it includes the five-pion interaction described by $\Gamma_0(U)$ [see Eq.~\eqref{eq:Gamma_0}] and the mediation of two spin-1 mesons, \ie, $a\to V^*V^*\to (2P)(2P)$. Although such decay channels are heavily suppressed by the four-body final-state phase space, large-width mesons such as $\rho^\pm$ and $\rho^0$, which mainly decay to pion pairs, can still make great contributions to the decay widths. Following Refs.~\cite{Aloni:2018vki,Ovchynnikov:2025gpx}, we explicitly calculated the decay widths of $a\to 2\pi^+2\pi^-$ and $a\to \pi^+\pi^-2\pi_0$ and further checked that at $m_a\sim 1.6~{\rm GeV}$ the values of the decay widths do smoothly connect to those of the on-shell decay modes $a\to 2\rho^0$ and $a\to \rho^+\rho^-$.

\subsection{$a\to 3P$ channels}

For such channels, we follow Ref.~\cite{Ovchynnikov:2025gpx} to incorporate the contact interactions arising from the four-pion interactions in the chiral Lagragian as well as the interactions mediated by off-shell scalar, vector, axial-vector, and tensor mesons. Note that different from Ref.~\cite{Ovchynnikov:2025gpx}, we do not consider the mixing between the axion and the axial-vector mesons, whose effects, as we have discussed in Section~\ref{subsec:mixing}, should be subdominant compared to the mixing between the axion and the pseudoscalar mesons.

\subsection{$a\to 2P1V$ channels}
Such channels are closely related to the $a\to 4P$ channels except that one of the intermediate spin-1 particles goes on shell instead of decaying promptly. 

\subsection{Combined results}\label{subsec:plots}

\begin{figure}[ht!]
    \centering
    \includegraphics[width=0.99\linewidth]{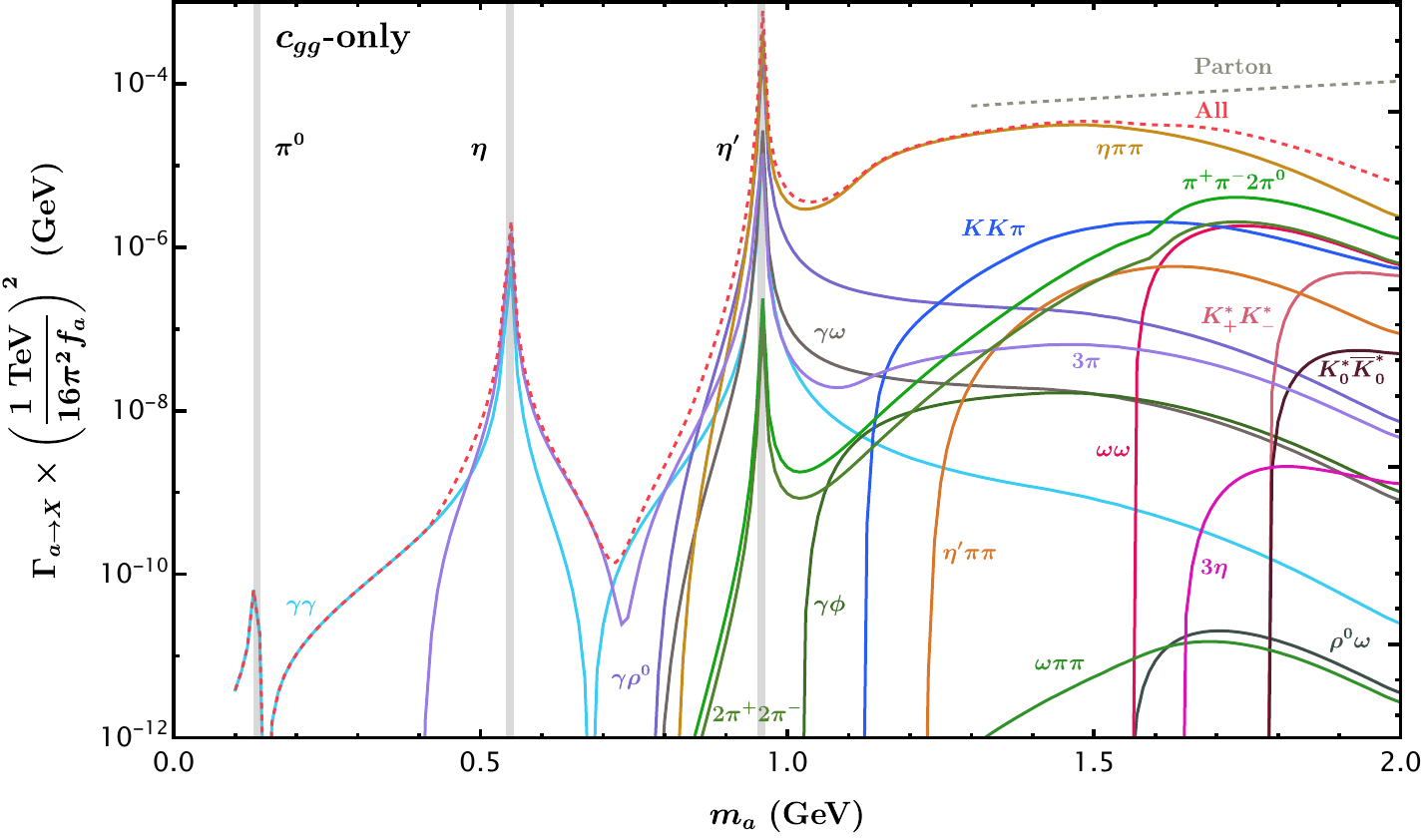}
    \caption{The axion decay widths against $m_a$ for the $c_{gg}$-only model with $c_{gg}=1$. The three gray bands with the labels $\pi^0$, $\eta$, and $\eta^\prime$ denote the mass poles of the three pseudoscalar mesons, respectively. Note that ``All'' represents the sum of all the exclusive partial widths considered in the plot and ``Parton'' the inclusive decay width approximated by $\Gamma(a\to gg/q\bar{q})$.
    }
    \label{fig:cgg}
\end{figure}

\begin{figure}[ht!]
    \centering
    \includegraphics[width=0.99\linewidth]{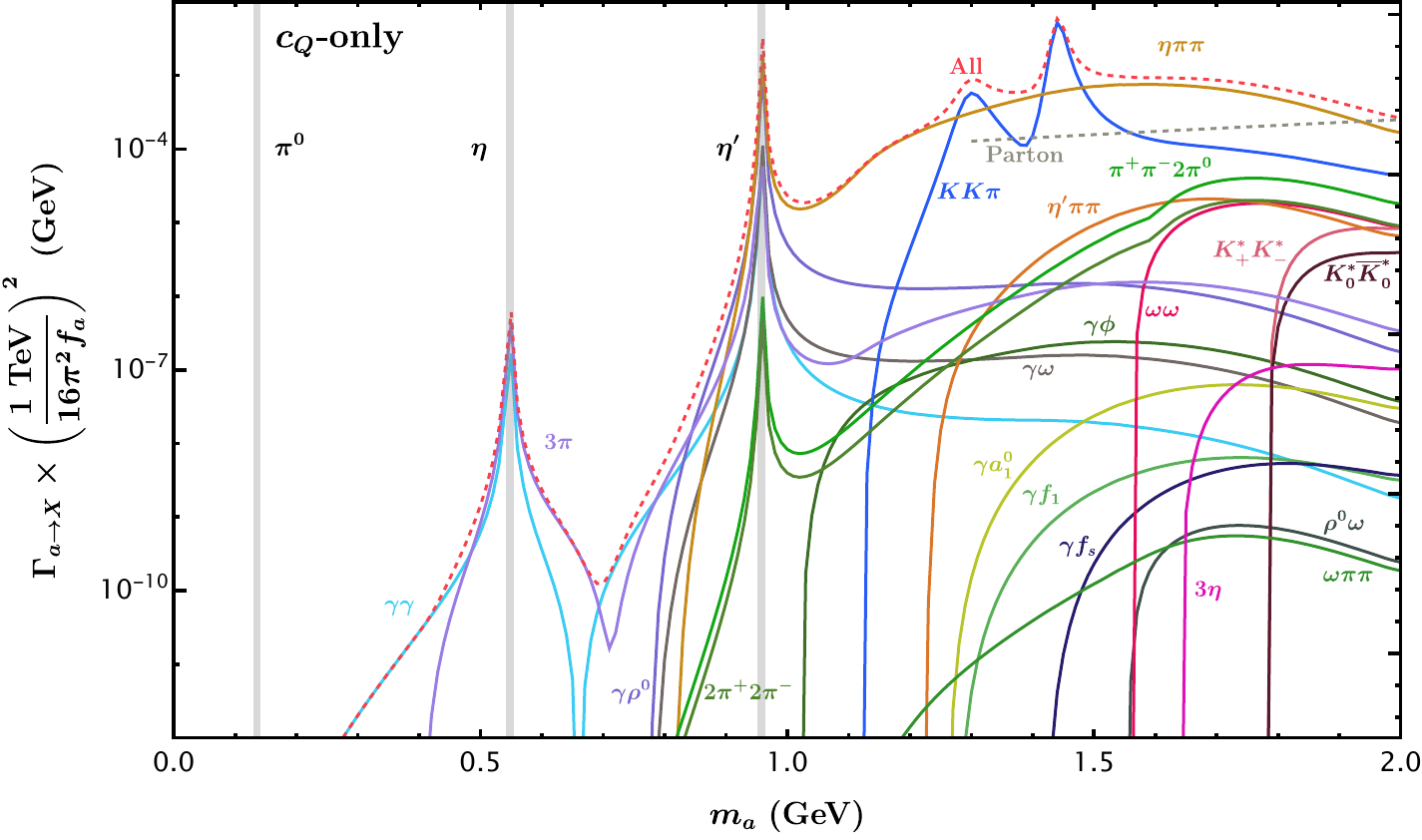}
    \caption{The same as Figure~\ref{fig:cgg} but for the $c_{Q}$-only model with $\clu=\cld=\cls=1$.
    The peaks above 1 GeV are from the mixing of heavy pseudo-scalar particle resonance peaks, see Ref.~\cite{Ovchynnikov:2025gpx} in detail.
    }
    \label{fig:cQ}
\end{figure}

\begin{figure}[ht!]
    \centering
    \includegraphics[width=0.99\linewidth]{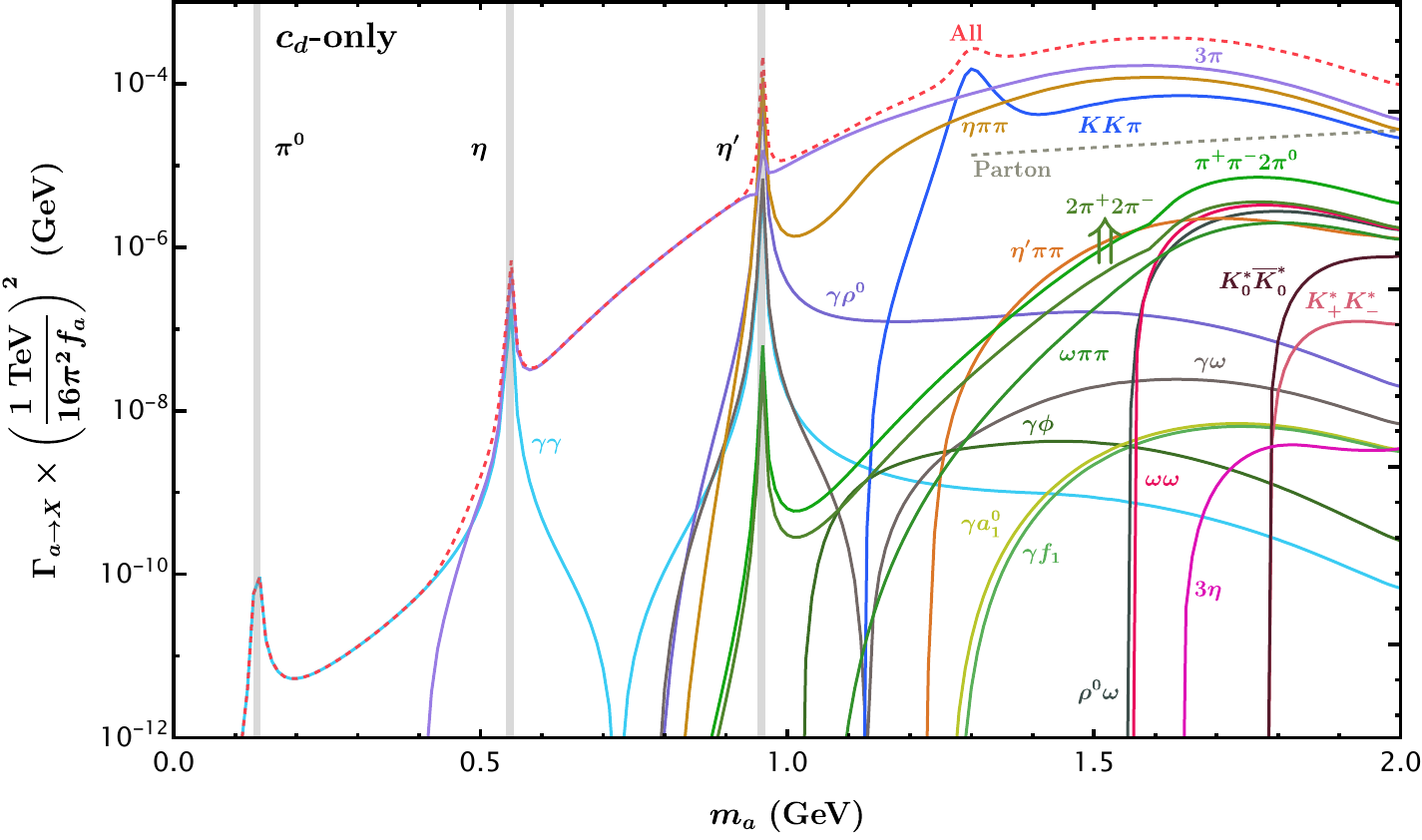}
    \caption{The same as Figure~\ref{fig:cgg} but for the $c_{d}$-only model with $\crd=1$. The peaks above 1 GeV are axion-heavy pseudo-scalar particle resonance peaks, see Ref.~\cite{Ovchynnikov:2025gpx} in detail. Note that, compared to Fig.~\ref{fig:cQ}, there is no higher-mass bump around 1.44 GeV due to $c_s = 0$.
    }
    \label{fig:cd}
\end{figure}

We consider the axion decays of three benchmark models: (1) the $\cgg$-only model with $\cgg=1$, (2) the $\cQ$-only model with $\clu=\cld=\cls\equiv c_Q=1$, and (3) the $c_d$-only model with $\crd\equiv c_d=1$, while all other axion couplings are set to zero in the individual models. Note that these three benchmark models represent three different interesting scenarios: (1) no axion-quark coupling at the quark-level, (2) isospin-symmetry conserved while C-symmmetry is violated, and (3) both isospin- and C-symmetries are broken. The corresponding partial widths of all the channels introduced previously predicted by these models are presented in Figures~\ref{fig:cgg}, \ref{fig:cQ}, and \ref{fig:cd}, respectively. For comparison, we also show the sum of all exclusive partial widths considered (labeled as ``All'') as well as the inclusive decay width approximated by $\Gamma(a\to gg/q\bar{q})$ (labeled as ``Parton''), for each model in the plots. One can see that they do not match well for all models, potentially owing to the ambiguity in describing the GeV-scale physics with either perturbative QCD or chiral Lagrangian. We also show the $\pi^0$, $\eta$, and $\eta^\prime$ mass poles with the corresponding gray bands in the plots. 

In the light-$m_a$ regime, the $2\gamma$ mode is the only kinematically allowed mode, followed by the $3\pi$ mode once $m_a\geq3m_\pi$. After that, three additional modes will be activated once $m_a$ reaches their respective thresholds: $\gamma\rho^0$, $\gamma\omega$, and $\eta\pi\pi$, all of which have partial widths comparable to the previous two modes except in the $c_d$-only model, where the $3\pi$ decay width is significantly larger than the others. The $2\pi^+2\pi^-$ and $\pi^+\pi^-2\pi^0$ modes follow shortly, but are both relatively suppressed at this point and will only become more significant when $m_a\geq 2m_\rho$ since these two modes are mediated by two $\rho$-mesons. Once $m_a\gtrsim1$~GeV, several other modes will be activated in the order of their respective kinematic thresholds, the most significant being the $\eta\pi\pi$, $KK\pi$, $2\pi^+2\pi^-$, $\pi^+\pi^-2\pi^0$, and $\eta^\prime\pi\pi$ modes, with the exception that the $3\pi$ mode is again the most dominant in the $c_d$-only model. We remark that for the $c_Q$- and $c_d$-only models, the $KK\pi$ mode also receives contributions from the heavy pseudoscalar mesons $\eta(1295)$ and $\eta(1440)$. These contributions are described by the Extended Linear Sigma Model~\cite{Parganlija:2016yxq} and implemented in Ref.~\cite{Ovchynnikov:2025gpx}; however, certain limitations of the model have still yet been addressed, and thus the numerical results should be taken with reservation (see Ref.~\cite{Ovchynnikov:2025gpx} for details).

Notice that the C-odd $\gamma f_1$ and $\gamma a_1^0$ modes are absent from the $\cgg$-only model, while the $\gamma f_s$ mode, which is also C-odd, is activated only in the $c_Q$-only model since $c_{f_s\gamma}^{\rm eff}\propto Q_d(c_s^L+c_s^R)$ and is only non-zero in this model for the three models considered here. One can also see that the $2V$ decay modes (except $2\gamma$ in the low-mass regime) are always subdominant, the most significant of which is the $\rho^*(2\pi)\rho^*(2\pi)$ mode, and thus probing the WZW interactions via axion decays is expected to be challenging. Nonetheless, one can still consider the phenomenological consequences of the WZW interactions without dominant $2V$ decays, such as the production of $\omega+a$ induced by the $a\omega\gamma$ interaction at low-energy colliders, which we have studied in Ref.~\cite{Bai:2024lpq}. 

\section{Discussion and conclusions}\label{sec:conclusions}

In this work, going beyond our previous work with two-flavor quarks, we have derived the complete three-flavor axion-interaction Lagrangian below the QCD scale, including the standard chiral Lagrangian and the full WZW term. With the QCD $U(1)_A$ anomaly and instanton effects properly incorporated, we have demonstrated that the independence of the auxiliary phases of the $aVV$ interactions is automatically achieved in all chirally related bases without any requirement on $\Tr(\boldsymbol{\kappa}_q)$, a feature not achieved in the previous literature. Compared to the literature, our formalism can also correctly describe both the C-even and C-odd interactions that involve both vector and axial-vector fields, which further demands the non-trivial cancellation of the $\boldsymbol{\delta}_q$ phases. Using three benchmark axion-interaction models, we calculated the decay patterns of the axion and discussed the relevant phenomenology to the WZW interactions studied in this work.

As we briefly discussed in Section~\ref{subsec:plots}, although the branching ratios of the $2V$ decay modes are mostly subdominant, one can still explore the possible phenomenology induced by the axion WZW interactions without relying on the associated decay channels. For instance, the production of axions at electron-ion colliders could be induced by the $a\omega\gamma$ interaction. Moreover, as we have studied in Ref.~\cite{Bai:2024lpq} based on the two-flavor framework, the associated production of $a+\omega$ at low-energy colliders such as BESIII and Super-Tau-Charm facility could also be induced by the same coupling. Compared to this previous work, in which we have truncated the calculation at $m_a=m_\eta$ as restricted by the two-flavor framework, one can in principle extend the same analysis up to higher axion masses using the three-flavor framework presented in this study. Moreover, if C-odd interactions are allowed in the axion WZW sector, one can further consider the associated production of $a+f_1$, with $f_1$ predominantly decaying to either $4\pi$ or $\eta\pi\pi$. These show that in addition to the consistent theoretical framework, the axion WZW term derived in this study further offers various opportunities for phenomenological studies.

\vspace{1cm}
\subsubsection*{Note added}
We thank Ferruccio Feruglio \textit{et al.}~\cite{Feruglio:2025xvc} for pointing out a typo in the axion-photon effective coefficient in an earlier version of this manuscript. We note that all our calculations were performed within a coherent numerical framework using our public code \href{https://github.com/nun3366/Axion-WZW-3}{\color{black}{\faGithub\,\underline{nun3366/Axion-WZW-3}}}, which yields self-consistent results in agreement with existing literature. Consequently, our physical results and conclusions remain unaffected by this textual error.

\vspace{1cm}
\subsubsection*{Acknowledgments}
We thank Maksym Ovchynnikov for helpful discussion. Y.B. is supported by the U.S. Department of Energy under the contract DE-SC-0017647 and DEAC02-06CH11357 at Argonne National Laboratory. 
T.K.C. is also supported by the Ministry of Education, Taiwan, under the Government Scholarship to Study Abroad.
The work of J.L. is supported by the National Science Foundation of China under Grant No. 12235001 and No. 12475103.
J.L. also thank APCTP, Pohang, Korea, for their hospitality during the focus program [APCTP-2025-F01], from which this work greatly benefited. J.L. and X.L.M. gratefully acknowledge the valuable discussions and insights provided by the members of the Collaboration of Precision Testing and New Physics.

\setlength{\bibsep}{3pt}
\bibliographystyle{JHEP}
\bibliography{references}

@article{Feruglio:2025xvc,
  author        = {Feruglio, Ferruccio and Levati, Gabriele and Ziegler, Robert},
  title         = {{On the decay of a light spinless particle into two photons}},
  eprint        = {2512.20695},
  archiveprefix = {arXiv},
  primaryclass  = {hep-ph},
  month         = {12},
  year          = {2025}
}

@article{Kawarabayashi:1980dp,
    author = "Kawarabayashi, Ken and Ohta, Nobuyoshi",
    title = "{The Problem of $\eta$ in the Large $N$ Limit: Effective Lagrangian Approach}",
    reportNumber = "UT-KOMABA-80-7",
    doi = "10.1016/0550-3213(80)90024-3",
    journal = "Nucl. Phys. B",
    volume = "175",
    pages = "477--492",
    year = "1980"
}

@article{Kawarabayashi:1980uh,
    author = "Kawarabayashi, Ken and Ohta, Nobuyoshi",
    title = "{On the Partial Conservation of the U(1) Current}",
    reportNumber = "UT-KOMABA-80-15",
    doi = "10.1143/PTP.66.1789",
    journal = "Prog. Theor. Phys.",
    volume = "66",
    pages = "1789",
    year = "1981"
}

@article{Cheng:2021kjg,
    author = "Cheng, Hsin-Chia and Li, Lingfeng and Salvioni, Ennio",
    title = "{A theory of dark pions}",
    eprint = "2110.10691",
    archivePrefix = "arXiv",
    primaryClass = "hep-ph",
    reportNumber = "CERN-TH-2021-150",
    doi = "10.1007/JHEP01(2022)122",
    journal = "JHEP",
    volume = "01",
    pages = "122",
    year = "2022"
}

@article{Goudzovski:2022vbt,
    author = "Goudzovski, Evgueni and others",
    title = "{New physics searches at kaon and hyperon factories}",
    eprint = "2201.07805",
    archivePrefix = "arXiv",
    primaryClass = "hep-ph",
    reportNumber = "FERMILAB-PUB-22-057-T",
    doi = "10.1088/1361-6633/ac9cee",
    journal = "Rept. Prog. Phys.",
    volume = "86",
    number = "1",
    pages = "016201",
    year = "2023"
}

@article{Harvey:2007rd,
    author = "Harvey, Jeffrey A. and Hill, Christopher T. and Hill, Richard J.",
    title = "{Anomaly mediated neutrino-photon interactions at finite baryon density}",
    eprint = "0708.1281",
    archivePrefix = "arXiv",
    primaryClass = "hep-ph",
    reportNumber = "EFI-PREPRINT-07-21, FERMILAB-PUB-07-401-T",
    doi = "10.1103/PhysRevLett.99.261601",
    journal = "Phys. Rev. Lett.",
    volume = "99",
    pages = "261601",
    year = "2007"
}

@article{Harvey:2007ca,
    author = "Harvey, Jeffrey A. and Hill, Christopher T. and Hill, Richard J.",
    title = "{Standard Model Gauging of the Wess-Zumino-Witten Term: Anomalies, Global Currents and pseudo-Chern-Simons Interactions}",
    eprint = "0712.1230",
    archivePrefix = "arXiv",
    primaryClass = "hep-th",
    reportNumber = "EFI-07-27, FERMILAB-PUB-07-628-T",
    doi = "10.1103/PhysRevD.77.085017",
    journal = "Phys. Rev. D",
    volume = "77",
    pages = "085017",
    year = "2008"
}

@article{Coloma:2023oxx,
    author = "Coloma, Pilar and Mart\'\i{}n-Albo, Justo and Urrea, Salvador",
    title = "{Discovering long-lived particles at DUNE}",
    eprint = "2309.06492",
    archivePrefix = "arXiv",
    primaryClass = "hep-ph",
    reportNumber = "IFT-UAM/CSIC-23-111, IFIC/23-40, FTUV-23-0823.4331",
    doi = "10.1103/PhysRevD.109.035013",
    journal = "Phys. Rev. D",
    volume = "109",
    number = "3",
    pages = "035013",
    year = "2024"
}

@article{HIKE:2023ext,
    author = "Ashraf, M. U. and others",
    collaboration = "HIKE",
    title = "{High Intensity Kaon Experiments (HIKE) at the CERN SPS Proposal for Phases 1 and 2}",
    eprint = "2311.08231",
    archivePrefix = "arXiv",
    primaryClass = "hep-ex",
    reportNumber = "CERN-SPSC-2023-031",
    month = "11",
    year = "2023"
}

@article{ICARUS:2024oqb,
    author = "Alrahman, F. Abd and others",
    collaboration = "ICARUS",
    title = "{Search for a Hidden Sector Scalar from Kaon Decay in the Dimuon Final State at ICARUS}",
    eprint = "2411.02727",
    archivePrefix = "arXiv",
    primaryClass = "hep-ex",
    reportNumber = "FERMILAB-PUB-24-0581-PPD",
    doi = "10.1103/PhysRevLett.134.151801",
    journal = "Phys. Rev. Lett.",
    volume = "134",
    number = "15",
    pages = "151801",
    year = "2025"
}

@article{BaBar:2021ich,
    author = "Lees, J. P. and others",
    collaboration = "BaBar",
    title = "{Search for an Axionlike Particle in $B$ Meson Decays}",
    eprint = "2111.01800",
    archivePrefix = "arXiv",
    primaryClass = "hep-ex",
    reportNumber = "BABAR-PUB-21/006, SLAC-PUB-17631",
    doi = "10.1103/PhysRevLett.128.131802",
    journal = "Phys. Rev. Lett.",
    volume = "128",
    number = "13",
    pages = "131802",
    year = "2022"
}

@article{Gao:2024vkw,
    author = "Gao, Rui and Hao, Jin and Duan, Chun-Gui and Guo, Zhi-Hui and Oller, J. A. and Zhou, Hai-Qing",
    title = "{Isospin-breaking contribution to the model-independent axion-photon-photon coupling in U(3) chiral theory}",
    eprint = "2411.06737",
    archivePrefix = "arXiv",
    primaryClass = "hep-ph",
    doi = "10.1140/epjc/s10052-025-13807-9",
    journal = "Eur. Phys. J. C",
    volume = "85",
    number = "1",
    pages = "97",
    year = "2025"
}

@article{Guo:2025icf,
    author = "Guo, Qian-Qian and Cao, Xiong-Hui and Guo, Zhi-Hui and Zhou, Hai-Qing",
    title = "{Axion production from electron-nucleon scattering in chiral effective theory}",
    eprint = "2504.06786",
    archivePrefix = "arXiv",
    primaryClass = "hep-ph",
    month = "4",
    year = "2025"
}

@article{Alves:2024dpa,
    author = "Alves, Daniele S. M. and Gonz\`alez-Sol\'\i{}s, Sergi",
    title = "{Final state rescattering effects in axio-hadronic \ensuremath{\eta} and \ensuremath{\eta}' decays}",
    eprint = "2402.02993",
    archivePrefix = "arXiv",
    primaryClass = "hep-ph",
    reportNumber = "LA-UR-24-20793",
    doi = "10.1007/JHEP07(2024)264",
    journal = "JHEP",
    volume = "07",
    pages = "264",
    year = "2024"
}

@article{Witten:1979vv,
    author = "Witten, Edward",
    title = "{Current Algebra Theorems for the U(1) Goldstone Boson}",
    reportNumber = "HUTP-79/A014",
    doi = "10.1016/0550-3213(79)90031-2",
    journal = "Nucl. Phys. B",
    volume = "156",
    pages = "269--283",
    year = "1979"
}

@article{Wang:2024tre,
    author = "Wang, Jin-Bao and Guo, Zhi-Hui and Lu, Zhun and Zhou, Hai-Qing",
    title = "{Axion production in the \ensuremath{\eta} \textrightarrow{} \ensuremath{\pi}\ensuremath{\pi}a decay within SU(3) chiral perturbation theory}",
    eprint = "2403.16064",
    archivePrefix = "arXiv",
    primaryClass = "hep-ph",
    doi = "10.1007/JHEP11(2024)029",
    journal = "JHEP",
    volume = "11",
    pages = "029",
    year = "2024"
}

@article{Georgi:1986df,
    author = "Georgi, Howard and Kaplan, David B. and Randall, Lisa",
    title = "{Manifesting the Invisible Axion at Low-energies}",
    reportNumber = "HUTP-86/A004",
    doi = "10.1016/0370-2693(86)90688-X",
    journal = "Phys. Lett. B",
    volume = "169",
    pages = "73--78",
    year = "1986"
}

@article{Gan:2020aco,
    author = "Gan, Liping and Kubis, Bastian and Passemar, Emilie and Tulin, Sean",
    title = "{Precision tests of fundamental physics with \ensuremath{\eta} and \ensuremath{\eta}' mesons}",
    eprint = "2007.00664",
    archivePrefix = "arXiv",
    primaryClass = "hep-ph",
    reportNumber = "JLAB-THY-20-3219",
    doi = "10.1016/j.physrep.2021.11.001",
    journal = "Phys. Rept.",
    volume = "945",
    pages = "1--105",
    year = "2022"
}

@article{Witten:1983tw,
    author = "Witten, Edward",
    title = "{Global Aspects of Current Algebra}",
    reportNumber = "PRINT-83-0262 (PRINCETON)",
    doi = "10.1016/0550-3213(83)90063-9",
    journal = "Nucl. Phys. B",
    volume = "223",
    pages = "422--432",
    year = "1983"
}

@article{tHooft:1979rat,
    author = "'t Hooft, Gerard",
    editor = "'t Hooft, Gerard and Itzykson, C. and Jaffe, A. and Lehmann, H. and Mitter, P. K. and Singer, I. M. and Stora, R.",
    title = "{Naturalness, chiral symmetry, and spontaneous chiral symmetry breaking}",
    reportNumber = "PRINT-80-0083 (UTRECHT)",
    doi = "10.1007/978-1-4684-7571-5_9",
    journal = "NATO Sci. Ser. B",
    volume = "59",
    pages = "135--157",
    year = "1980"
}

@article{DiVecchia:2013swa,
    author = "Di Vecchia, Paolo and Sannino, Francesco",
    title = "{The Physics of the $\theta$-angle for Composite Extensions of the Standard Model}",
    eprint = "1310.0954",
    archivePrefix = "arXiv",
    primaryClass = "hep-ph",
    reportNumber = "CP3-ORIGINS-2013-34, DIAS-2013-34",
    doi = "10.1140/epjp/i2014-14262-4",
    journal = "Eur. Phys. J. Plus",
    volume = "129",
    pages = "262",
    year = "2014"
}

@article{DiVecchia:1980yfw,
    author = "Di Vecchia, P. and Veneziano, G.",
    title = "{Chiral Dynamics in the Large n Limit}",
    reportNumber = "CERN-TH-2814",
    doi = "10.1016/0550-3213(80)90370-3",
    journal = "Nucl. Phys. B",
    volume = "171",
    pages = "253--272",
    year = "1980"
}

@article{Witten:1980sp,
    author = "Witten, Edward",
    title = "{Large N Chiral Dynamics}",
    reportNumber = "HUTP-80/A005",
    doi = "10.1016/0003-4916(80)90325-5",
    journal = "Annals Phys.",
    volume = "128",
    pages = "363",
    year = "1980"
}

@article{Bai:2023bbg,
    author = "Bai, Yang and de Lima, Carlos Henrique",
    title = "{Electrobaryonic axion: hair of neutron stars}",
    eprint = "2311.18794",
    archivePrefix = "arXiv",
    primaryClass = "hep-ph",
    doi = "10.1007/JHEP05(2024)312",
    journal = "JHEP",
    volume = "05",
    pages = "312",
    year = "2024"
}

@article{Wess:1971yu,
    author = "Wess, J. and Zumino, B.",
    title = "{Consequences of anomalous Ward identities}",
    doi = "10.1016/0370-2693(71)90582-X",
    journal = "Phys. Lett. B",
    volume = "37",
    pages = "95--97",
    year = "1971"
}

@article{Svrcek:2006yi,
    author = "Svrcek, Peter and Witten, Edward",
    title = "{Axions In String Theory}",
    eprint = "hep-th/0605206",
    archivePrefix = "arXiv",
    reportNumber = "SLAC-PUB-11894",
    doi = "10.1088/1126-6708/2006/06/051",
    journal = "JHEP",
    volume = "06",
    pages = "051",
    year = "2006"
}

@article{Aloni:2018vki,
    author = "Aloni, Daniel and Soreq, Yotam and Williams, Mike",
    title = "{Coupling QCD-Scale Axionlike Particles to Gluons}",
    eprint = "1811.03474",
    archivePrefix = "arXiv",
    primaryClass = "hep-ph",
    reportNumber = "CERN-TH-2018-237, MIT-CTP/5080, MIT-CTP-5080",
    doi = "10.1103/PhysRevLett.123.031803",
    journal = "Phys. Rev. Lett.",
    volume = "123",
    number = "3",
    pages = "031803",
    year = "2019"
}

@article{ParticleDataGroup:2022pth,
    author = "Workman, R. L. and others",
    collaboration = "Particle Data Group",
    title = "{Review of Particle Physics}",
    doi = "10.1093/ptep/ptac097",
    journal = "PTEP",
    volume = "2022",
    pages = "083C01",
    year = "2022"
}

@article{Harada:2003jx,
    author = "Harada, Masayasu and Yamawaki, Koichi",
    title = "{Hidden local symmetry at loop: A New perspective of composite gauge boson and chiral phase transition}",
    eprint = "hep-ph/0302103",
    archivePrefix = "arXiv",
    reportNumber = "DPNU-03-02",
    doi = "10.1016/S0370-1573(03)00139-X",
    journal = "Phys. Rept.",
    volume = "381",
    pages = "1--233",
    year = "2003"
}

@article{Kaymakcalan:1983qq,
    author = "Kaymakcalan, O. and Rajeev, S. and Schechter, J.",
    title = "{Nonabelian Anomaly and Vector Meson Decays}",
    reportNumber = "SU-4222-278, COO-3533-278",
    doi = "10.1103/PhysRevD.30.594",
    journal = "Phys. Rev. D",
    volume = "30",
    pages = "594",
    year = "1984"
}

@article{Son:2004tq,
    author = "Son, D. T. and Zhitnitsky, Ariel R.",
    title = "{Quantum anomalies in dense matter}",
    eprint = "hep-ph/0405216",
    archivePrefix = "arXiv",
    reportNumber = "INT-PUB-04-13",
    doi = "10.1103/PhysRevD.70.074018",
    journal = "Phys. Rev. D",
    volume = "70",
    pages = "074018",
    year = "2004"
}

@article{Peccei:1977hh,
    author = "Peccei, R. D. and Quinn, Helen R.",
    title = "{CP Conservation in the Presence of Instantons}",
    reportNumber = "ITP-568-STANFORD",
    doi = "10.1103/PhysRevLett.38.1440",
    journal = "Phys. Rev. Lett.",
    volume = "38",
    pages = "1440--1443",
    year = "1977"
}

@article{Peccei:1977ur,
    author = "Peccei, R. D. and Quinn, Helen R.",
    title = "{Constraints Imposed by CP Conservation in the Presence of Instantons}",
    reportNumber = "ITP-572-STANFORD",
    doi = "10.1103/PhysRevD.16.1791",
    journal = "Phys. Rev. D",
    volume = "16",
    pages = "1791--1797",
    year = "1977"
}

@article{Weinberg:1977ma,
    author = "Weinberg, Steven",
    title = "{A New Light Boson?}",
    reportNumber = "HUTP-77/A074",
    doi = "10.1103/PhysRevLett.40.223",
    journal = "Phys. Rev. Lett.",
    volume = "40",
    pages = "223--226",
    year = "1978"
}

@article{Wilczek:1977pj,
    author = "Wilczek, Frank",
    title = "{Problem of Strong  $P$  and  $T$  Invariance in the Presence of Instantons}",
    reportNumber = "Print-77-0939 (COLUMBIA)",
    doi = "10.1103/PhysRevLett.40.279",
    journal = "Phys. Rev. Lett.",
    volume = "40",
    pages = "279--282",
    year = "1978"
}

@article{Bauer:2021wjo,
    author = "Bauer, Martin and Neubert, Matthias and Renner, Sophie and Schnubel, Marvin and Thamm, Andrea",
    title = "{Consistent Treatment of Axions in the Weak Chiral Lagrangian}",
    eprint = "2102.13112",
    archivePrefix = "arXiv",
    primaryClass = "hep-ph",
    reportNumber = "IPPP/20-82, MITP/21-007, ZU-TH-01/21",
    doi = "10.1103/PhysRevLett.127.081803",
    journal = "Phys. Rev. Lett.",
    volume = "127",
    number = "8",
    pages = "081803",
    year = "2021"
}

@article{Bando:1987br,
    author = "Bando, Masako and Kugo, Taichiro and Yamawaki, Koichi",
    title = "{Nonlinear Realization and Hidden Local Symmetries}",
    reportNumber = "DPNU-87-63, AICHI-1, KUNS-903",
    doi = "10.1016/0370-1573(88)90019-1",
    journal = "Phys. Rept.",
    volume = "164",
    pages = "217--314",
    year = "1988"
}

@article{Bauer:2020jbp,
    author = "Bauer, Martin and Neubert, Matthias and Renner, Sophie and Schnubel, Marvin and Thamm, Andrea",
    title = "{The Low-Energy Effective Theory of Axions and ALPs}",
    eprint = "2012.12272",
    archivePrefix = "arXiv",
    primaryClass = "hep-ph",
    reportNumber = "IPPP/20/69, MITP/20-070 SISSA 30/2020/FISI, ZH-TH-47/20",
    doi = "10.1007/JHEP04(2021)063",
    journal = "JHEP",
    volume = "04",
    pages = "063",
    year = "2021"
}

@article{Bauer:2021mvw,
    author = "Bauer, Martin and Neubert, Matthias and Renner, Sophie and Schnubel, Marvin and Thamm, Andrea",
    title = "{Flavor probes of axion-like particles}",
    eprint = "2110.10698",
    archivePrefix = "arXiv",
    primaryClass = "hep-ph",
    reportNumber = "MITP/21-025, CERN-TH-2021-148, IPPP/21/37",
    doi = "10.1007/JHEP09(2022)056",
    journal = "JHEP",
    volume = "09",
    pages = "056",
    year = "2022"
}

@article{Bauer:2017ris,
    author = "Bauer, Martin and Neubert, Matthias and Thamm, Andrea",
    title = "{Collider Probes of Axion-Like Particles}",
    eprint = "1708.00443",
    archivePrefix = "arXiv",
    primaryClass = "hep-ph",
    reportNumber = "MITP-17-047",
    doi = "10.1007/JHEP12(2017)044",
    journal = "JHEP",
    volume = "12",
    pages = "044",
    year = "2017"
}

@article{GrillidiCortona:2015jxo,
    author = "Grilli di Cortona, Giovanni and Hardy, Edward and Pardo Vega, Javier and Villadoro, Giovanni",
    title = "{The QCD axion, precisely}",
    eprint = "1511.02867",
    archivePrefix = "arXiv",
    primaryClass = "hep-ph",
    doi = "10.1007/JHEP01(2016)034",
    journal = "JHEP",
    volume = "01",
    pages = "034",
    year = "2016"
}

@article{Blinov:2021say,
    author = "Blinov, Nikita and Kowalczyk, Elizabeth and Wynne, Margaret",
    title = "{Axion-like particle searches at DarkQuest}",
    eprint = "2112.09814",
    archivePrefix = "arXiv",
    primaryClass = "hep-ph",
    reportNumber = "FERMILAB-PUB-21-749-V",
    doi = "10.1007/JHEP02(2022)036",
    journal = "JHEP",
    volume = "02",
    pages = "036",
    year = "2022"
}

@article{Kaiser:2000gs,
    author = "Kaiser, Roland and Leutwyler, H.",
    title = "{Large N(c) in chiral perturbation theory}",
    eprint = "hep-ph/0007101",
    archivePrefix = "arXiv",
    reportNumber = "BUTP-00-19",
    doi = "10.1007/s100520000499",
    journal = "Eur. Phys. J. C",
    volume = "17",
    pages = "623--649",
    year = "2000"
}

@article{tHooft:1986ooh,
    author = "'t Hooft, Gerard",
    title = "{How Instantons Solve the U(1) Problem}",
    reportNumber = "PRINT-86-0358 (UTRECHT)",
    doi = "10.1016/0370-1573(86)90117-1",
    journal = "Phys. Rept.",
    volume = "142",
    pages = "357--387",
    year = "1986"
}

@article{Bai:2024lpq,
    author = "Bai, Yang and Chen, Ting-Kuo and Liu, Jia and Ma, Xiaolin",
    title = "{Wess-Zumino-Witten Interactions of Axions}",
    eprint = "2406.11948",
    archivePrefix = "arXiv",
    primaryClass = "hep-ph",
    doi = "10.1103/PhysRevLett.134.081803",
    journal = "Phys. Rev. Lett.",
    volume = "134",
    number = "8",
    pages = "081803",
    year = "2025"
}

@article{DHoker:1984izu,
    author = "D'Hoker, Eric and Farhi, Edward",
    title = "{Decoupling a Fermion Whose Mass Is Generated by a Yukawa Coupling: The General Case}",
    reportNumber = "MIT-CTP-1165",
    doi = "10.1016/0550-3213(84)90586-8",
    journal = "Nucl. Phys. B",
    volume = "248",
    pages = "59--76",
    year = "1984"
}

@article{DHoker:1984mif,
    author = "D'Hoker, Eric and Farhi, Edward",
    title = "{Decoupling a Fermion in the Standard Electroweak Theory}",
    reportNumber = "MIT-CTP-1166",
    doi = "10.1016/0550-3213(84)90587-X",
    journal = "Nucl. Phys. B",
    volume = "248",
    pages = "77",
    year = "1984"
}

@article{Ovchynnikov:2025gpx,
    author = "Ovchynnikov, Maksym and Zaporozhchenko, Andrii",
    title = "{ALPs coupled to gluons in the GeV mass range -- data-driven and consistent}",
    eprint = "2501.04525",
    archivePrefix = "arXiv",
    primaryClass = "hep-ph",
    reportNumber = "CERN-TH-2025-006",
    month = "1",
    year = "2025"
}

@article{Veneziano:1979ec,
    author = "Veneziano, G.",
    title = "{U(1) Without Instantons}",
    reportNumber = "CERN-TH-2651",
    doi = "10.1016/0550-3213(79)90332-8",
    journal = "Nucl. Phys. B",
    volume = "159",
    pages = "213--224",
    year = "1979"
}

@article{Dekens:2022gha,
    author = "Dekens, Wouter and de Vries, Jordy and Shain, Sachin",
    title = "{CP-violating axion interactions in effective field theory}",
    eprint = "2203.11230",
    archivePrefix = "arXiv",
    primaryClass = "hep-ph",
    doi = "10.1007/JHEP07(2022)014",
    journal = "JHEP",
    volume = "07",
    pages = "014",
    year = "2022"
}

@article{Geng:2025fuc,
    author = "Geng, Chao-Qiang and Liu, Chia-Wei and Wu, Yue-Liang",
    title = "{Identify hadron anomalous couplings at colliders}",
    eprint = "2504.14979",
    archivePrefix = "arXiv",
    primaryClass = "hep-ph",
    month = "4",
    year = "2025"
}

@article{Rosenzweig:1979ay,
    author = "Rosenzweig, C. and Schechter, J. and Trahern, C. G.",
    editor = "Brezin, E. and Wadia, S. R.",
    title = "{Is the Effective Lagrangian for QCD a Sigma Model?}",
    reportNumber = "SU-4217-148, COO-3533-148",
    doi = "10.1103/PhysRevD.21.3388",
    journal = "Phys. Rev. D",
    volume = "21",
    pages = "3388",
    year = "1980"
}

@article{Bando:1984ej,
    author = "Bando, M. and Kugo, T. and Uehara, S. and Yamawaki, K. and Yanagida, T.",
    title = "{Is rho Meson a Dynamical Gauge Boson of Hidden Local Symmetry?}",
    reportNumber = "RRK 84-22",
    doi = "10.1103/PhysRevLett.54.1215",
    journal = "Phys. Rev. Lett.",
    volume = "54",
    pages = "1215",
    year = "1985"
}

@article{Fujiwara:1984mp,
    author = "Fujiwara, Takanori and Kugo, Taichiro and Terao, Haruhiko and Uehara, Shozo and Yamawaki, Koichi",
    title = "{Nonabelian Anomaly and Vector Mesons as Dynamical Gauge Bosons of Hidden Local Symmetries}",
    reportNumber = "KUNS-764",
    doi = "10.1143/PTP.73.926",
    journal = "Prog. Theor. Phys.",
    volume = "73",
    pages = "926",
    year = "1985"
}

@article{BaBar:2004ytv,
    author = "Aubert, Bernard and others",
    collaboration = "BaBar",
    title = "{Study of $e^+e^- \to \pi^+ \pi^- \pi^0$ process using initial state radiation with BaBar}",
    eprint = "hep-ex/0408078",
    archivePrefix = "arXiv",
    reportNumber = "SLAC-PUB-10624, BABAR-PUB-04-034",
    doi = "10.1103/PhysRevD.70.072004",
    journal = "Phys. Rev. D",
    volume = "70",
    pages = "072004",
    year = "2004"
}

@article{BaBar:2007ceh,
    author = "Aubert, Bernard and others",
    collaboration = "BaBar",
    title = "{Measurements of $e^{+} e^{-} \to K^{+} K^{-} \eta$, $K^{+} K^{-} \pi^0$ and $K^0_{s} K^\pm \pi^\mp$ cross- sections using initial state radiation events}",
    eprint = "0710.4451",
    archivePrefix = "arXiv",
    primaryClass = "hep-ex",
    reportNumber = "SLAC-PUB-12968, BABAR-PUB-07-052",
    doi = "10.1103/PhysRevD.77.092002",
    journal = "Phys. Rev. D",
    volume = "77",
    pages = "092002",
    year = "2008"
}

@article{BaBar:2017zmc,
    author = "Lees, J. P. and others",
    collaboration = "BaBar",
    title = "{Measurement of the ${e}^{+}{e}^{{-}}{\rightarrow}{{\pi}}^{+}{{\pi}}^{{-}}{{\pi}}^{0}{{\pi}}^{0}$ cross section using initial-state radiation at BABAR}",
    eprint = "1709.01171",
    archivePrefix = "arXiv",
    primaryClass = "hep-ex",
    reportNumber = "SLAC-PUB-17147, BABAR-PUB-17-002",
    doi = "10.1103/PhysRevD.96.092009",
    journal = "Phys. Rev. D",
    volume = "96",
    number = "9",
    pages = "092009",
    year = "2017"
}

@article{Kaiser:1990yf,
    author = "Kaiser, Norbert and Meissner, Ulf G.",
    title = "{Generalized hidden symmetry for low-energy hadron physics}",
    reportNumber = "BUTP-90-22-BERN, NBI-90-27",
    doi = "10.1016/0375-9474(90)90431-K",
    journal = "Nucl. Phys. A",
    volume = "519",
    pages = "671--696",
    year = "1990"
}

@article{Parganlija:2016yxq,
    author = "Parganlija, Denis and Giacosa, Francesco",
    title = "{Excited Scalar and Pseudoscalar Mesons in the Extended Linear Sigma Model}",
    eprint = "1612.09218",
    archivePrefix = "arXiv",
    primaryClass = "hep-ph",
    doi = "10.1140/epjc/s10052-017-4962-y",
    journal = "Eur. Phys. J. C",
    volume = "77",
    number = "7",
    pages = "450",
    year = "2017"
}

\end{document}